\documentclass{aa520}
\usepackage{txfonts}
\usepackage{graphicx}
\usepackage{natbib}
\bibpunct{(}{)}{;}{a}{}{,}

\newcommand{\UX}    {UX~Ori}
\newcommand{\kms}   {km\,s$^{-1}$}
\newcommand{\ms}    {m\,s$^{-2}$}
\newcommand{\Dv}    {$\Delta v$}

\newcommand{\Hb}    {H$\beta$}
\newcommand{\Hg}    {H$\gamma$}
\newcommand{\Hd}    {H$\delta$}

\newcommand{\Hz}    {H$\zeta$}

\newcommand{\Rstar} {$R_\star$}

\newcommand{\Msun}  {$M_\odot$}

\begin{document}
\title{A dynamical study of the circumstellar gas in UX Orionis}

\author{A.~Mora         \inst{1}        \and
        A.~Natta        \inst{2}        \and
        C.~Eiroa        \inst{1}        \and
        C.A.~Grady      \inst{3}        \and
        D.~de Winter    \inst{4}        \and
        J.K.~Davies     \inst{5}        \and
        R.~Ferlet       \inst{6}        \and
        A.W.~Harris     \inst{7}        \and
        B.~Montesinos   \inst{8,9}        \and
        R.D.~Oudmaijer  \inst{10}       \and
        J.~Palacios     \inst{1}        \and
        A.~Quirrenbach  \inst{11}       \and
        H.~Rauer        \inst{7}        \and
        A.~Alberdi      \inst{8}        \and
        A.~Cameron      \inst{12}       \and
        H.J.~Deeg       \inst{13}       \and
        F.~Garz\'on     \inst{13}       \and
        K.~Horne        \inst{12}       \and
        B.~Mer\'{\i}n   \inst{9}        \and
        A.~Penny        \inst{14}       \and
        J.~Schneider    \inst{15}       \and
        E.~Solano       \inst{9}        \and
        Y.~Tsapras      \inst{12}       \and
        P.R.~Wesselius  \inst{16}}

\institute{
Departamento de F\'{\i}sica Te\'orica C-XI, Universidad Aut\'onoma de Madrid, Cantoblanco 28049 Madrid, Spain
        \and
Osservatorio Astrofisico di Arcetri, Largo Fermi 5, I-50125 Firenze, Italy
        \and
NOAO/STIS, Goddard Space Flight Center, Code 681, NASA/GSFC, Greenbelt, MD 20771, USA
        \and
TNO/TPD-Space Instrumentation, Stieltjesweg 1, PO Box 155, 2600 AD Delft, The Netherlands
        \and
Astronomy Technology Centre, Royal Observatory, Blackford Hill, Edinburgh, EH9 3HJ, UK
	\and
CNRS, Institute d'Astrophysique de Paris, 98bis Bd. Arago, 75014 Paris, France 
        \and
DLR Department of Planetary Exploration, Rutherfordstrasse 2, 12489 Berlin, Germany
        \and
Instituto de Astrof\'{\i}sica de Andaluc\'{\i}a, Apartado de Correos 3004, 18080 Granada, Spain
        \and
LAEFF, VILSPA, Apartado de Correos 50727, 28080 Madrid, Spain
        \and
Department of Physics and Astronomy, University of Leeds, Leeds LS2 9JT, UK
        \and
Department of Physics, Center for Astrophysics and Space Sciences, University of California San Diego, Mail Code 0424, La Jolla, CA 92093-0424, USA
        \and
Physics \& Astronomy, University of St. Andrews, North Haugh, St. Andrews KY16 9SS, Scotland, UK
        \and
Instituto de Astrof\'{\i}sica de Canarias, La Laguna 38200 Tenerife, Spain
        \and
Rutherford Appleton Laboratory, Didcot, Oxfordshire OX11 0QX, UK
        \and
Observatoire de Paris, 92195 Meudon, France
        \and
SRON, Universiteitscomplex ``Zernike'', Landleven 12, P.O. Box 800, 9700 AV Groningen, The Netherlands
}

\offprints{Alcione Mora,
\email{alcione.mora@uam.es}}

\date{Received 14 May 2002 / Accepted 11 July 2002}

\abstract{
We present the results of a high spectral resolution ($\lambda / \Delta \lambda $ = 49000) study of the circumstellar (CS) gas around the intermediate mass, pre-main sequence star \object{UX Ori}.
The results are based on a set of 10 \'echelle spectra covering the spectral range 3800 -- 5900 \AA, monitoring the star on time scales of months, days and hours.
A large number of transient blueshifted and redshifted absorption features are detected in the Balmer and in many  metallic lines.
A multigaussian fit is applied to determine for each transient absorption the velocity, $v$, dispersion velocity, $\Delta v$, and the parameter $R$, which provides a measure of the absorption strength of the CS gas.
The time evolution of those parameters is presented and discussed.
A comparison of intensity ratios among the transient absorptions suggests a solar-like composition of the CS gas.
This confirms previous results and excludes a very metal-rich environment as the cause of the transient features in UX Ori.
The features can be grouped by their similar velocities into 24 groups, of which 17 are redshifted and 7 blueshifted.
An analysis of the velocity of the groups allows us to identify them as signatures of the dynamical evolution of 7 clumps of gas, of which 4 represent accretion events and 3 outflow events.
Most of the events decelerate at a rate of tenths of m\,s$^{-2}$, while 2 events accelerate at approximately the same rate; one event is seen experiencing both an acceleration and a deceleration phase and lasts for a period of few days.
This time scale seems to be the typical duration of outflowing and infalling events in UX Ori. 
The dispersion velocity and the relative aborption strength of the features do not show drastic changes during the lifetime of the events, which suggests they are gaseous blobs preserving their geometrical and physical identity.
These data are a very useful tool for constraining and validating theoretical models of the chemical and physical conditions of CS gas around young stars; in particular, we suggest that the simultaneous presence of infalling and outflowing gas should be investigated in the context of detailed magnetospheric accretion models, similar to those proposed for the lower mass T Tauri stars.  
}

\titlerunning{A dynamical study of the circumstellar gas in UX Orionis}
\authorrunning{Mora et al.}

\maketitle


\section{Introduction}

The detection of planetesimals is highly relevant for the study of the formation and evolution of planetary systems, since it is nowadays accepted that planets form from CS disks via the formation of planetesimals \citep{beckwith2000}.
Several lines of evidence suggest that the young main sequence A5V star $\beta$~Pic \citep[20 Myr, ][]{barrado1999} hosts planetesimals inside its large CS disk.
The main argument is the presence of transient Redshifted Absorption Components (RACs) in high resolution spectra of strong metallic lines, like \ion{Ca}{ii}~K 3934~\AA.
These spectroscopic events have been interpreted as being caused by the evaporation of comet-like highly hydrogen-depleted bodies.
The interpretation is known as the Falling Evaporating Bodies (FEB) scenario \citep[ and references therein]{lagrange2000}.
However, the presence or absence of planetesimals during the pre-main sequence (PMS) phase is a controversial observational topic.
The time scale for the formation of planetesimals \citep[$\sim$10$^4$~yr, ][]{beckwith2000} is shorter than the duration of the PMS phase ($\sim$1--10~Myr), which suggests that they should exist during PMS stellar evolution.

UX~Ori-like PMS objects (UXORs) are characterized by a peculiar photo-polarimetric variability, which has been interpreted as the signature of massive, almost edge-on, CS disks \citep{grinin1991}.
Most UXORs have A spectral types and therefore are the PMS evolutionary precursors of $\beta$~Pic.  
UXORs have been reported to show RACs \citep{grinin1994,dewinter1999}; they also show Blueshifted Absorption Components (BACs) in their spectra. 
In this paper, the acronym TAC (Transient Absorption Component) will be used to denote both RACs and BACs.
In analogy to $\beta$~Pic, RACs observed in UXORs have been interpreted in terms of the FEB scenario \citep{grady2000}.
However, this interpretation has recently been questioned by \citet{natta2000}, who used the spectra obtained by \citet{grinin2001} to analyze the dynamics and chemical composition of a very strong, redshifted event in UX Ori itself, an A4 IV star \citep{mora2001} $\sim 2\times 10^6$ year old \citep{natta1999}.
Gas accretion from a CS disk was suggested in \citet{natta2000} as an alternative to the FEB scenario to explain the observed RACs in UX Ori.
In addition, \citet{beust2001} have found that the FEB hypothesis cannot produce detectable transient absorptions in typical HAe CS conditions, unless the stars are relatively old ($\ge 10^7$ yr).

A detailed observational study of the kinematics and chemistry of TACs, e. g. by means of \'echelle spectra which simultaneously record many metallic and hydrogen lines, can discriminate between the two scenarios. 
For instance, in a FEB event large metallicities are expected, while gas accreted from a CS disk would have approximately solar abundances.
A strong observational requirement is set by the time scale of monitoring of the TACs.
In \citet{natta2000} spectra were taken 3 days apart.
Since UX Ori is a highly variable star, there is some ambiguity in identifying transient spectral components of different velocities detected over this time interval as having the same physical origin.
A better time resolution is needed in order to ensure that the TACs observed at different velocities  are due to the dynamical evolution of the same gas.
The EXPORT collaboration \citep{eiroa2000a} obtained high resolution \'echelle spectra of a large sample of PMS stars \citep{mora2001}.
About 10 PMS stars showed TACs which were intensively monitored ($\Delta$t~$\leq$~1~day).
The study of the kinematics and chemistry of these events provides important tools for identifying their origin, or at least to put severe constraints on it.   

This paper presents an analysis of the TACs observed in UX Ori by EXPORT and shows that the results are not compatible with a FEB scenario.
The layout of the paper is as follows:
Section 2 presents a brief description of the EXPORT observations.
Section 3 presents the procedures followed in the analysis of the spectra.
Section 4 presents the kinematic and chemical results of the detected TACs.
Section 5 gives a discussion of the dynamics and nature of the gas.
In section 6, we present our conclusions.

\section{Observations}

High resolution  spectra of \UX\ were taken in October 1998 and January 1999 using the Utrecht Echelle Spectrograph (UES) at the 4.2m WHT (La Palma Observatory).
We collected 10 \'echelle spectra in the wavelength range 3800~--~5900~\AA, with resolution $\lambda / \Delta\lambda$~=~49000 (6~km/s); exposure times range between 20 and 30 minutes.
The observing log is given in Table~\ref{photometry}, in which the long-term (months) and short-term (days, hours) monitoring is evident.
Standard MIDAS and IRAF procedures have been used for the spectroscopic reduction \citep{mora2001}.
Final typical signal to noise ratio (SNR) values are $\sim$150.

\begin{table}
\caption{EXPORT UES/WHT UX~Ori observing log. The Julian date (-2450000) of each  exposure is given in column 1.
Columns 2 to 5 give simultaneous $V$, $H$, $K$ and \%P$_V$ photopolarimetric data, when available, taken from \citet{oudmaijer2001} and \citet{eiroa2001}.}
\label{photometry}
\centerline{
\begin{tabular}{lllll}
\hline
\hline
Julian date & $V$    & \%P$_V$ & $H$  & $K$  \\
\hline
1112.5800   &  9.94  &  1.33   & 8.19 & 7.37 \\
1113.6034   &  9.82  &  1.32   & 8.20 & 7.37 \\
1113.7194   &  --    &  --     & --   & --   \\
\hline
1207.5268   &  --    &  --     & --   & --   \\
1208.5072   &  --    &  --     & 8.41 & 7.56 \\
1209.4204   &  9.80  &  1.15   & 8.31 & 7.62 \\
1210.3317   &  9.89  &  1.26   & --   & --   \\
1210.3568   &  9.86  &  1.34   & --   & --   \\
1210.4237   &  9.88  &  1.30   & 8.34 & --   \\
1210.5156   &  9.90  &  1.31   & 8.34 & --   \\
\hline
\end{tabular}}
\end{table}

EXPORT also carried out simultaneous intermediate resolution spectroscopy, optical photo-polarimetry and near-IR photometry of UX Ori during the nights when the UES spectra were taken \citep{mora2001,oudmaijer2001,eiroa2001}.
Table~\ref{photometry} also gives photo-polarimetric data taken simultaneously with the UES spectra; they show that UX Ori was always in its bright state with little variation, but significant \citep{oudmaijer2001}.

\begin{figure}
\centerline{\includegraphics[height=\hsize,angle=-90,clip=true]{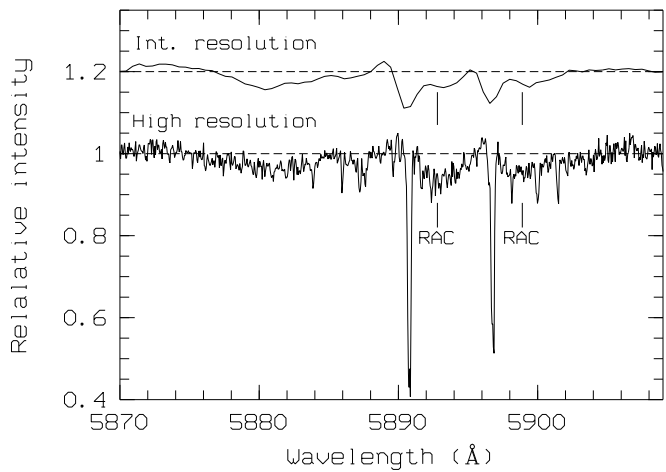}}
\caption{Comparison between intermediate (top, J.D. 2451210.3098, 10~min. exposure time) and high (bottom, J.D. 2451210.3317, 30~min. exposure time) resolution spectra.
The \ion{Na}{i}~D spectral region is shown.
A sharp IS zero velocity component and a much broader RAC (indicated in the figure) can be seen in both lines of the doublet in the high resolution spectrum.
They are also detected in the intermediate resolution spectrum, but the big difference in the width of the components is not clear.
For the sake of clarity  an artificial vertical displacement of 0.2 units has been applied to the intermediate resolution spectrum.}
\label{wht_vs_int}
\end{figure}

\section{Analysis of the spectra}

A detailed analysis of TACs requires high  SNR ($>$~25) and high spectral resolution ($<$~10~km/s) in order to resolve the kinematic components observed simultaneously in different lines, as is shown in Fig. 1.
This figure presents the \ion{Na}{i}~D lines in  one of the UES spectra together with the simultaneous intermediate resolution spectrum (R~$\sim$~6000) in the same lines. 
The UES spectrum clearly shows a sharp, IS, narrow zero velocity component and a RAC in both lines of the doublet while both components, though visible, cannot be cleanly separated in the lower resolution spectrum.
This is why we restrict our further analysis  of the absorptions to the UES spectra.

All the \UX\ UES spectra show CS spectral features in a variety of lines.
The features are seen in absorption and can be either blueshifted or redshifted; some underlying emission is also present (see below).
Their profiles are complex and blended components are directly seen in many cases.  
Thus, the analysis of the CS contribution to the observed spectra requires a careful subtraction of the stellar photospheric spectrum and a method to characterize the blended kinematical components.
In this section we describe the procedure we have followed, which is the one used in \citet{natta2000}, although we have improved it by considering all detected TACs simultaneously.

\subsection{Subtraction of the UX Ori photospheric spectrum}

Firstly we estimate the radial velocity of the star.
The heliocentric correction of each observed spectrum is computed using MIDAS.
Then, the radial velocity of the star is estimated by fitting the  Na~I~D sharp IS absorption.
This absorption  is  stable in velocity (velocity differences in our spectra are less than 1.0~km/s) and is related to the  radial velocity of the star \citep{finkenzeller1984}.
A value of 18.3~$\pm$~1.0~km/s is obtained, in good agreement with the 18~km/s estimate by \citet{grinin1994}.   

Kurucz (1993) model atmospheres assuming solar metallicity and turbulence velocity of 2~km/s have been used to synthesize the photospheric spectra, which are later used for a comparison with the observed ones.
Atomic line data have been obtained from the VALD online database \citep{kupka1999}, and the codes SYNTHE \citep{kurucz1979,jeffery1996} and SYNSPEC \citep{hubeny1995} are used to synthesize the metallic lines and Balmer hydrogen lines, respectively. 
In this way we compute a large grid of synthetic spectra with the effective temperature, gravity and rotation velocity as free variable parameters. 

\begin{figure}
\centerline{\includegraphics[height=\hsize,angle=-90,clip=true]{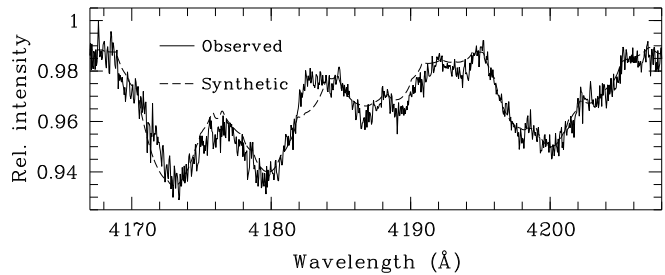}}
\caption{Comparison of the observed average spectrum of UX Ori (solid line) to the synthetic spectrum with stellar parameters T$_{\rm eff}$~=~9250~K, log~g~=~4.0 and ${\it v\sin i}$~=~215~km/s (dashed line).  
The spectral region shown includes the blends at 4172~--~4179~{\AA} (mainly \ion{Fe}{ii} and \ion{Ti}{ii}) and 4203~{\AA} (\ion{Fe}{i}).}
\label{obs_vs_syn}
\end{figure}

The best set of free parameters is estimated by comparing a number of lines among the observed spectra and the synthetic ones.
This is not an easy  task since most UX Ori lines are variable to some extent, and the choice of pure photospheric lines is not trivial.
Our choice is to consider a number of weak lines that show a minimum degree of variability (less that 1\%) in the UES spectra.
The blends located at 4172~--~4179~{\AA} (mainly \ion{Fe}{ii} and \ion{Ti}{ii}) and 4203~\AA\ (\ion{Fe}{i}) have been used to estimate T$_{\rm eff}$ (effective temperature) and log~g (logarithm of the surface gravity), since they are sensitive to changes in T$_{\rm eff}$ and log~g for early type A stars \citep{gray1987,gray1989b}.
The synthetic spectra have also been broadened allowing for stellar rotation.
The best set of free parameters we find is T$_{\rm eff}$~=~9250~K, log~g~=~4.0, {\it v $\sin i$}~=~215~km/s.  
Fig.~\ref{obs_vs_syn} shows the comparison between the synthetic and the average observed spectrum in the above-mentioned lines; the agreement is excellent.
The stellar parameters of the synthetic spectrum are in very good agreement with the spectral type and rotational velocity derived elsewhere \citep{mora2001}.

\begin{figure}
\centerline{\includegraphics[width=\hsize,clip=true]{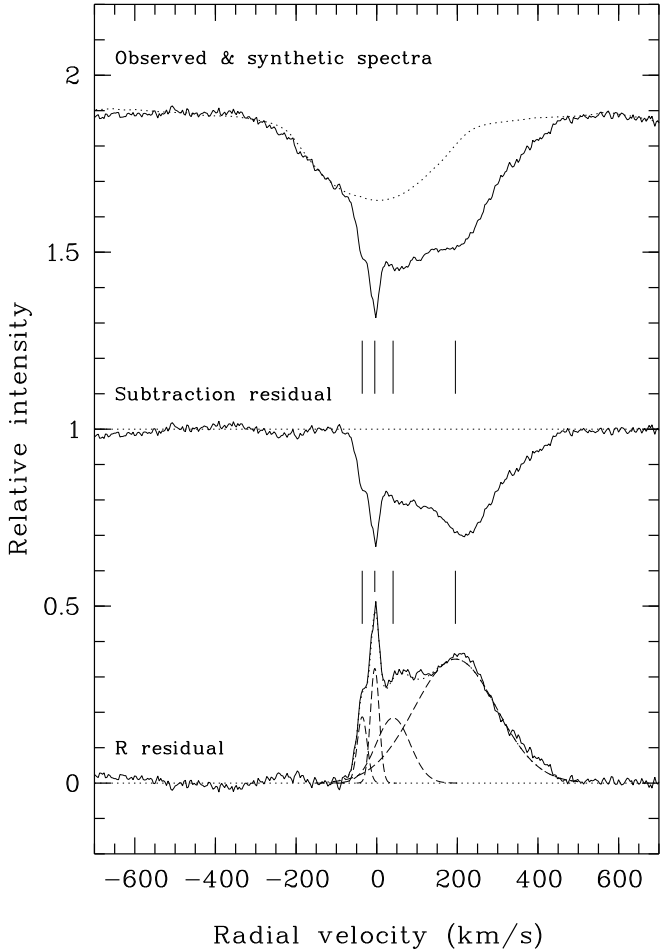}}
\caption{Subtraction of the photospheric spectrum. 
The spectral range corresponds to the \ion{Ca}{ii}~K line. 
Top: observed (solid line) and synthetic (dotted line) spectra are displayed.
Middle: Residual due to the CS contribution (observed -- photospheric).
Bottom: Normalized residual absorption, $R$ = (1 -- observed~/~photospheric).
Vertical tick marks indicate the 4 absorption components of the \ion{Ca}{ii}~K which can be identified in this spectrum, including the narrow IS absorption.
In the bottom we show the 4 gaussians (dashed lines), whose sum (dotted line) excellently reproduces the $R$ profile (solid line).} 
\label{subtraction}
\end{figure}

The next step consists of the subtraction of the best synthetic photospheric spectrum from each observed spectrum. 
This has been carried out by fitting the continuum with a linear law between two points selected at both sides of those lines showing TACs.
As an example Fig.~\ref{subtraction} shows the \ion{Ca}{ii}~K line results, where the synthetic spectrum and the J.D. 2451209.4204 UX Ori observed one are plotted together with the residual after the subtraction.
The  residual corresponds to the CS contribution to the observed spectra.
As a more convenient way to represent this contribution we define the $R$ parameter (normalized residual absorption):

\centerline{$R = 1 - F_{\rm obs} / F_{\rm syn}$} 

$R$~=~0 means no CS absorption, and $R$~=~1 denotes complete stellar light occultation, i.e. $R$ quantifies the CS absorption strength and could be directly compared to the models in 
\citet{natta2000}.
$R$ is also plotted in Fig.~\ref{subtraction}. 
  
\subsection{R multicomponent gaussian fitting}

The CS contribution presents a very complex line profile caused by the blending of different transient features. 
As an example, Fig.~\ref{subtraction} shows the components of the \ion{Ca}{ii}~K line which are clearly present in that particular observation. 
For the sake of analysing the profile we assume that it is caused by gas with different kinematics and that each component can be represented by means of a gaussian function.
This choice provides a direct interpretation of the fit parameters in terms of a physical characterization of the gaseous transient absorption components: the gaussian center gives the central velocity (V) of each component, the gaussian  FWHM (Full Width at Half Maximum) provides the velocity dispersion ($\Delta v$) and finally the gaussian peak gives the value of the residual absorption $R$.
Since the feaures are blended a multigaussian fit is needed, which has been carried out by means of the IRAF {\tt ngaussfit} routine. 
Fig.~\ref{subtraction} shows the results of the multigaussian fit for the \ion{Ca}{ii}~K line.
4 gaussians, i.e. 4 kinematical components, each with their corresponding parameters, provide an excellent agreement with the residual CS contribution profile.

This analysis is applied to the lines with the strongest variability, which are  good tracers of the CS gas properties: H$\beta$ 4861~\AA, H$\gamma$ 4340~\AA, H$\delta$ 4102~\AA, H$\epsilon$ 3970~\AA, H$\zeta$ 3889~\AA, \ion{Ca}{ii}~K 3934~\AA, \ion{Ca}{ii}~H 3968~\AA, \ion{Na}{i}~D2 5890~{\AA} and \ion{Na}{i}~D1 5896~\AA. 
All Balmer lines with good SNR  present transient absorption features (even the noisy H$\kappa$~3750~\AA, 10$^{\rm th}$ line in the Balmer series). 
They are also observed in many other lines (e.g. \ion{He}{i}, \ion{Fe}{i} and \ion{Fe}{ii}), as already pointed out by \citet{grinin2001}, but a corresponding discussion is deferred to a future work.

\begin{figure}
\centerline{\includegraphics[width=\hsize,clip=true]{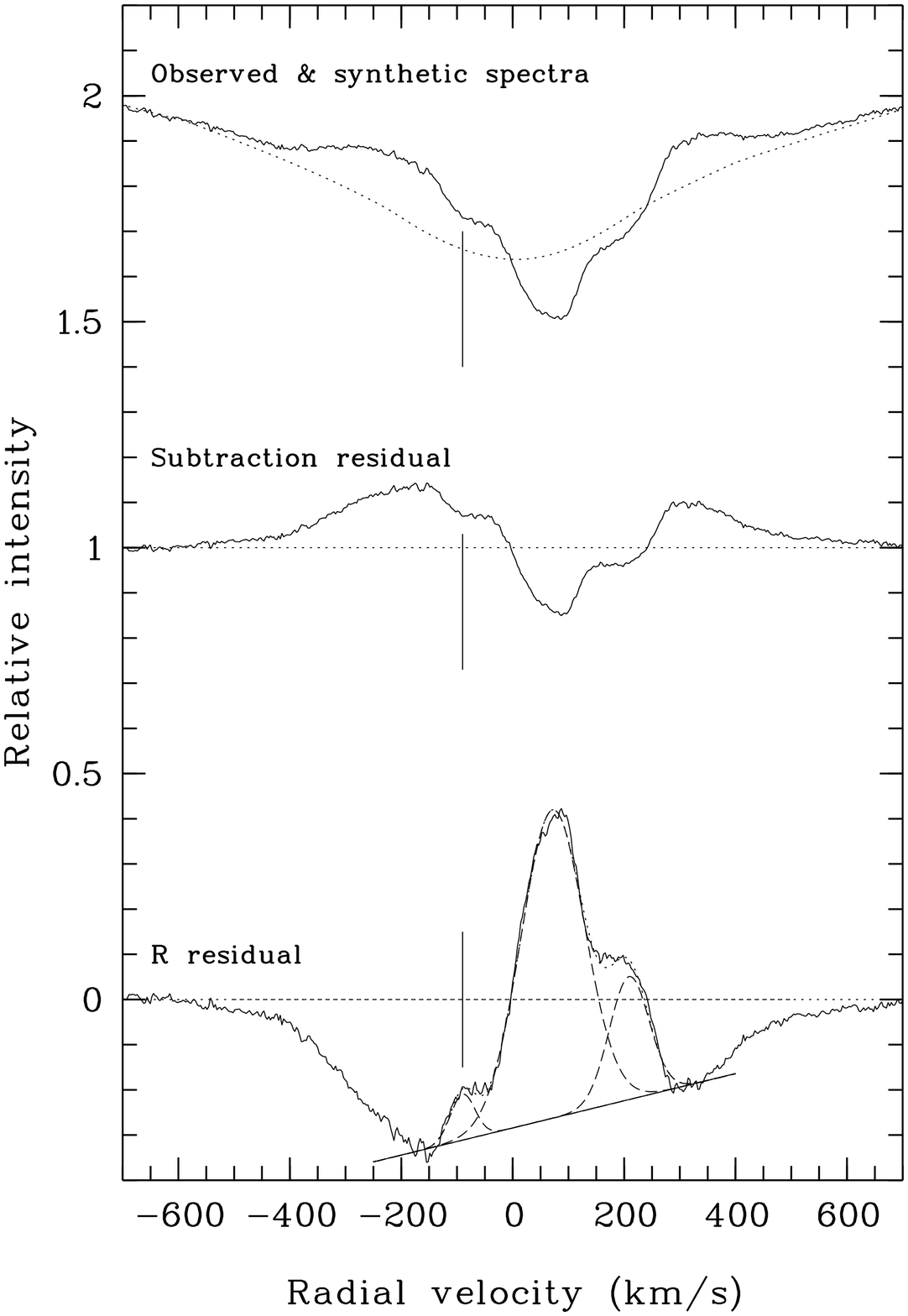}}
\caption{
Underlying emission in the H$\beta$ absorption line.
Top: Normalized observed spectrum (solid line) and photospheric model (dotted line). 
The line profile is clearly modulated by an emission contribution at $\simeq\pm$300~km/s in the line wings.
Middle: Subtracted spectrum (observed -- synthetic).
Bottom: $R$ curve. 
The vertical mark points to an absorption component which would be undetectable if the continuum level were not redefined.
This has been done by means of a linear fit to the peaks of the emission, shown as throughs in the $R$ plot.
This fit (solid line at the bottom) is the new zero level for the gaussians.
3 gaussians (dashed lines) whose sum (dotted line) provides the optimal fit to the $R$ curve are also plotted at the bottom.}
\label{hbemission}
\end{figure}

Underlying line emission is often significant in the Balmer lines, particularly in H$\beta$, H$\gamma$ and H$\delta$; in fact, H$\alpha$ is seen in emission (equivalent width $\sim$~10~\AA) in the simultaneous INT spectra.
\ion{Ca}{ii}~K and H also show little emission.
In these cases $R$ is underestimated and negative values can artificially arise when the synthetic photospheric spectrum is subtracted.
This is especially bad for faint absorption components near the emission peaks since the latter appear as minima in the $R$ plot.  
In order to avoid contamination by these emission components, we define a new ``zero level", estimated by a linear fit to the minima observed in the $R$ plot. Fig.~\ref{hbemission} illustrates this approach.
The consistency of the results obtained with different lines confirms its validity.  
Further improvements should be based on theoretical modelling or higher order fitting of the emission.

\begin{table*}
\caption{Identified transient absorption components in the Balmer and 
\ion{Ca}{ii} and \ion{Na}{i} lines of UX Ori.
JD is the Julian Date ($-$2450000).
Numbers in column 3 represent the event assigned to the particular absorptions (see text section 4).
A ``0'' in that column corresponds to the narrow IS absorptions, while a ``--" means that the absorption is not associated with a particular event. 
Columns 4 to 6 give: $v$, the radial velocity of the transient absorption, $\Delta v$, the FWHM and $R$,  the normalized residual absorption.}
\label{master_table}
\centerline{
\tiny
\begin{tabular}{lllrrl}
\hline
\hline
Line & JD & Event & $v$ (km/s) & $\Delta v$ (km/s) & $R$ \\
\hline
H$\beta$      & 1112.5800 & 1 & 209 &  84 & 0.27 \\
H$\gamma$     & 1112.5800 & 1 & 207 &  78 & 0.21 \\
H$\delta$     & 1112.5800 & 1 & 202 &  73 & 0.16 \\
H$\epsilon$   & 1112.5800 & 1 & 206 &  68 & 0.19 \\
\ion{Ca}{ii} K& 1112.5800 & 1 & 199 &  91 & 0.26 \\
\ion{Na}{i} D2& 1112.5800 & 1 & 169 & 132 & 0.10 \\
\ion{Na}{i} D1& 1112.5800 & 1 & 183 & 123 & 0.09 \\
H$\beta$      & 1112.5800 & 2 &  77 & 135 & 0.68 \\
H$\gamma$     & 1112.5800 & 2 &  63 & 172 & 0.59 \\
H$\delta$     & 1112.5800 & 2 &  72 & 183 & 0.43 \\
H$\epsilon$   & 1112.5800 & 2 &  56 & 199 & 0.55 \\
\ion{Ca}{ii} K& 1112.5800 & 2 &  62 & 155 & 0.49 \\
H$\beta$      & 1112.5800 & 3 &--90 &  52 & 0.10 \\
H$\gamma$     & 1112.5800 & 3 &--91 &  51 & 0.10 \\
H$\delta$     & 1112.5800 & 3 &--88 &  41 & 0.06 \\
H$\epsilon$   & 1112.5800 & 3 &--85 &  47 & 0.18 \\
\ion{Ca}{ii} K& 1112.5800 & 3 &--92 &  61 & 0.11 \\
\ion{Ca}{ii} K& 1112.5800 & 0 &   5 &  17 & 0.23 \\
\ion{Ca}{ii} H& 1112.5800 & 0 &   6 &  12 & 0.23 \\
\ion{Na}{i} D2& 1112.5800 & 0 &   6 &  11 & 0.65 \\
\ion{Na}{i} D1& 1112.5800 & 0 &   6 &  11 & 0.50 \\
H$\zeta$      & 1112.5800 & --& 111 & 225 & 0.25 \\
H$\beta$      & 1113.6034 & 1 & 132 & 238 & 0.41 \\
H$\gamma$     & 1113.6034 & 1 & 109 & 217 & 0.33 \\
H$\delta$     & 1113.6034 & 1 & 175 & 136 & 0.16 \\
H$\epsilon$   & 1113.6034 & 1 & 157 & 100 & 0.22 \\
H$\zeta$      & 1113.6034 & 1 & 104 & 258 & 0.18 \\
\ion{Ca}{ii} K& 1113.6034 & 1 & 114 & 246 & 0.31 \\
H$\beta$      & 1113.6034 & 2 &  55 &  75 & 0.58 \\
H$\gamma$     & 1113.6034 & 2 &  57 &  64 & 0.50 \\
H$\delta$     & 1113.6034 & 2 &  55 & 109 & 0.53 \\
H$\epsilon$   & 1113.6034 & 2 &  50 & 105 & 0.56 \\
H$\zeta$      & 1113.6034 & 2 &  56 &  71 & 0.23 \\
\ion{Ca}{ii} K& 1113.6034 & 2 &  53 &  59 & 0.41 \\
\ion{Ca}{ii} H& 1113.6034 & 2 &  60 &  58 & 0.63 \\
\ion{Na}{i} D2& 1113.6034 & 2 &  61 &  38 & 0.09 \\
H$\beta$      & 1113.6034 & 3 &--51 &  42 & 0.07 \\
H$\gamma$     & 1113.6034 & 3 &--30 &  69 & 0.07 \\
\ion{Ca}{ii} K& 1113.6034 & 3 &--56 &  45 & 0.06 \\
\ion{Ca}{ii} K& 1113.6034 & 0 &   2 &  17 & 0.30 \\
\ion{Ca}{ii} H& 1113.6034 & 0 &   5 &  15 & 0.26 \\
\ion{Na}{i} D2& 1113.6034 & 0 &   5 &  12 & 0.68 \\
\ion{Na}{i} D1& 1113.6034 & 0 &   5 &  12 & 0.53 \\
H$\beta$      & 1113.7194 & 1 & 114 & 188 & 0.36 \\
H$\gamma$     & 1113.7194 & 1 &  90 & 169 & 0.36 \\
H$\delta$     & 1113.7194 & 1 & 182 &  89 & 0.11 \\
H$\epsilon$   & 1113.7194 & 1 & 143 & 121 & 0.20 \\
H$\zeta$      & 1113.7194 & 1 & 165 &  84 & 0.09 \\
\ion{Ca}{ii} K& 1113.7194 & 1 &  84 & 205 & 0.35 \\
H$\beta$      & 1113.7194 & 2 &  52 &  75 & 0.62 \\
H$\gamma$     & 1113.7194 & 2 &  57 &  58 & 0.44 \\
H$\delta$     & 1113.7194 & 2 &  59 & 121 & 0.56 \\
H$\epsilon$   & 1113.7194 & 2 &  50 &  96 & 0.51 \\
H$\zeta$      & 1113.7194 & 2 &  57 & 108 & 0.38 \\
\ion{Ca}{ii} K& 1113.7194 & 2 &  56 &  50 & 0.35 \\
\ion{Ca}{ii} H& 1113.7194 & 2 &  60 &  60 & 0.60 \\
\ion{Na}{i} D2& 1113.7194 & 2 &  61 &  44 & 0.10 \\
\ion{Na}{i} D1& 1113.7194 & 2 &  60 &  28 & 0.06 \\
H$\beta$      & 1113.7194 & 3 &--44 &  72 & 0.13 \\
H$\gamma$     & 1113.7194 & 3 &--22 &  99 & 0.13 \\
\ion{Ca}{ii} K& 1113.7194 & 3 &--58 &  52 & 0.08 \\
\ion{Ca}{ii} K& 1113.7194 & 0 &   6 &  19 & 0.31 \\
\ion{Ca}{ii} H& 1113.7194 & 0 &   5 &  19 & 0.20 \\
\ion{Na}{i} D2& 1113.7194 & 0 &   5 &  13 & 0.64 \\
\ion{Na}{i} D1& 1113.7194 & 0 &   5 &  12 & 0.49 \\
H$\beta$      & 1207.5268 & 4 &  78 & 221 & 0.26 \\
H$\gamma$     & 1207.5268 & 4 &  48 & 211 & 0.29 \\
H$\delta$     & 1207.5268 & 4 &  52 & 223 & 0.24 \\
H$\epsilon$   & 1207.5268 & 4 &  68 & 207 & 0.17 \\
H$\zeta$      & 1207.5268 & 4 &  83 & 128 & 0.16 \\
\ion{Ca}{ii} K& 1207.5268 & 4 &  51 & 269 & 0.32 \\
H$\beta$      & 1207.5268 & 5 &--60 &  88 & 0.61 \\
H$\gamma$     & 1207.5268 & 5 &--53 &  90 & 0.46 \\
H$\delta$     & 1207.5268 & 5 &--49 &  82 & 0.39 \\
H$\epsilon$   & 1207.5268 & 5 &--50 &  90 & 0.19 \\
H$\zeta$      & 1207.5268 & 5 &--43 &  92 & 0.30 \\
\ion{Ca}{ii} K& 1207.5268 & 5 &--59 &  70 & 0.32 \\
\ion{Ca}{ii} H& 1207.5268 & 5 &--56 &  63 & 0.38 \\
\ion{Na}{i} D2& 1207.5268 & 5 &--54 &  22 & 0.10 \\
\ion{Ca}{ii} K& 1207.5268 & 0 &   6 &  33 & 0.17 \\
\ion{Ca}{ii} H& 1207.5268 & 0 &   1 &  21 & 0.15 \\
\ion{Na}{i} D2& 1207.5268 & 0 &   3 &  13 & 0.59 \\
\ion{Na}{i} D1& 1207.5268 & 0 &   3 &  12 & 0.49 \\
H$\beta$      & 1207.5268 & --&  20 &  62 & 0.20 \\
H$\gamma$     & 1208.5072 & 4 &  89 & 285 & 0.20 \\
H$\delta$     & 1208.5072 & 4 &  84 & 268 & 0.18 \\
H$\epsilon$   & 1208.5072 & 4 & 109 & 294 & 0.14 \\
H$\zeta$      & 1208.5072 & 4 &  91 & 331 & 0.15 \\
\ion{Ca}{ii} K& 1208.5072 & 4 &  77 & 279 & 0.25 \\
H$\beta$      & 1208.5072 & 5 &--35 & 107 & 0.60 \\
\hline
\end{tabular}
\hspace{1.5cm}
\begin{tabular}{lllrrl}
\hline
\hline
Line & JD & Event & $v$ (km/s) & $\Delta v$ (km/s) & $R$ \\
\hline
H$\gamma$     & 1208.5072 & 5 &--34 &  85 & 0.47 \\
H$\delta$     & 1208.5072 & 5 &--44 &  81 & 0.36 \\
H$\epsilon$   & 1208.5072 & 5 &--37 &  52 & 0.15 \\
H$\zeta$      & 1208.5072 & 5 &--40 &  76 & 0.20 \\
\ion{Ca}{ii} K& 1208.5072 & 5 &--38 &  42 & 0.33 \\
\ion{Ca}{ii} H& 1208.5072 & 5 &--39 &  34 & 0.28 \\
\ion{Na}{i} D2& 1208.5072 & 5 &--34 &  18 & 0.18 \\
\ion{Na}{i} D1& 1208.5072 & 5 &--36 &  22 & 0.10 \\
H$\beta$      & 1208.5072 & 6 &--95 &  30 & 0.10 \\
H$\gamma$     & 1208.5072 & 6 &--91 &  29 & 0.12 \\
\ion{Ca}{ii} K& 1208.5072 & 6 &--82 &  41 & 0.22 \\
\ion{Ca}{ii} H& 1208.5072 & 6 &--83 &  42 & 0.19 \\
\ion{Ca}{ii} K& 1208.5072 & 0 &   1 &  26 & 0.26 \\
\ion{Ca}{ii} H& 1208.5072 & 0 & --1 &   8 & 0.14 \\
\ion{Na}{i} D2& 1208.5072 & 0 &   3 &  12 & 0.61 \\
\ion{Na}{i} D1& 1208.5072 & 0 &   3 &  13 & 0.47 \\
H$\beta$      & 1209.4204 & 4 & 207 & 309 & 0.43 \\
H$\gamma$     & 1209.4204 & 4 & 200 & 267 & 0.36 \\
H$\delta$     & 1209.4204 & 4 & 201 & 263 & 0.28 \\
H$\epsilon$   & 1209.4204 & 4 & 135 & 278 & 0.49 \\
H$\zeta$      & 1209.4204 & 4 & 191 & 313 & 0.21 \\
\ion{Ca}{ii} K& 1209.4204 & 4 & 195 & 240 & 0.35 \\
\ion{Na}{i} D2& 1209.4204 & 4 & 143 & 111 & 0.04 \\
H$\beta$      & 1209.4204 & 5 &--11 &  76 & 0.48 \\
H$\gamma$     & 1209.4204 & 5 &--22 &  53 & 0.30 \\
H$\delta$     & 1209.4204 & 5 &--14 &  54 & 0.16 \\
\ion{Ca}{ii} K& 1209.4204 & 5 &--36 &  29 & 0.19 \\
\ion{Ca}{ii} H& 1209.4204 & 5 &--32 &  26 & 0.12 \\
\ion{Ca}{ii} K& 1209.4204 & 0 & --5 &  26 & 0.32 \\
\ion{Ca}{ii} H& 1209.4204 & 0 & --4 &  17 & 0.20 \\
\ion{Na}{i} D2& 1209.4204 & 0 &   3 &  12 & 0.62 \\
\ion{Na}{i} D1& 1209.4204 & 0 &   4 &  12 & 0.50 \\
H$\gamma$     & 1209.4204 & --&  35 &  85 & 0.15 \\
\ion{Ca}{ii} K& 1209.4204 & --&  41 &  99 & 0.18 \\
H$\beta$      & 1210.3317 & 4 & 138 & 287 & 0.66 \\
H$\gamma$     & 1210.3317 & 4 & 138 & 255 & 0.54 \\
H$\delta$     & 1210.3317 & 4 & 142 & 270 & 0.43 \\
H$\epsilon$   & 1210.3317 & 4 & 110 & 277 & 0.64 \\
H$\zeta$      & 1210.3317 & 4 & 151 & 265 & 0.31 \\
\ion{Ca}{ii} K& 1210.3317 & 4 & 124 & 265 & 0.54 \\
\ion{Na}{i} D2& 1210.3317 & 4 & 115 & 137 & 0.06 \\
\ion{Na}{i} D1& 1210.3317 & 4 & 133 & 150 & 0.06 \\
H$\beta$      & 1210.3317 & 7 &   3 &  46 & 0.19 \\
H$\gamma$     & 1210.3317 & 7 &   9 &  39 & 0.10 \\
\ion{Ca}{ii} K& 1210.3317 & 0 & --3 &  19 & 0.20 \\
\ion{Ca}{ii} H& 1210.3317 & 0 &   1 &   8 & 0.16 \\
\ion{Na}{i} D2& 1210.3317 & 0 &   1 &  12 & 0.57 \\
\ion{Na}{i} D1& 1210.3317 & 0 &   2 &  13 & 0.46 \\
H$\beta$      & 1210.3568 & 4 & 132 & 287 & 0.64 \\
H$\gamma$     & 1210.3568 & 4 & 132 & 260 & 0.53 \\
H$\delta$     & 1210.3568 & 4 & 128 & 238 & 0.43 \\
H$\epsilon$   & 1210.3568 & 4 & 107 & 280 & 0.63 \\
H$\zeta$      & 1210.3568 & 4 & 140 & 249 & 0.29 \\
\ion{Ca}{ii} K& 1210.3568 & 4 & 119 & 296 & 0.52 \\
\ion{Na}{i} D2& 1210.3568 & 4 & 113 & 109 & 0.04 \\
\ion{Na}{i} D1& 1210.3568 & 4 & 142 & 127 & 0.04 \\
H$\beta$      & 1210.3568 & 7 &   6 &  41 & 0.18 \\
H$\gamma$     & 1210.3568 & 7 &  17 &  37 & 0.08 \\
\ion{Ca}{ii} K& 1210.3568 & 0 &   0 &  15 & 0.14 \\
\ion{Ca}{ii} H& 1210.3568 & 0 & --2 &  12 & 0.14 \\
\ion{Na}{i} D2& 1210.3568 & 0 &   2 &  13 & 0.60 \\
\ion{Na}{i} D1& 1210.3568 & 0 &   3 &  13 & 0.48 \\
H$\beta$      & 1210.4237 & 4 & 134 & 240 & 0.66 \\
H$\gamma$     & 1210.4237 & 4 & 122 & 242 & 0.55 \\
H$\delta$     & 1210.4237 & 4 & 125 & 229 & 0.44 \\
H$\epsilon$   & 1210.4237 & 4 &  89 & 253 & 0.64 \\
H$\zeta$      & 1210.4237 & 4 & 126 & 230 & 0.30 \\
\ion{Ca}{ii} K& 1210.4237 & 4 & 117 & 272 & 0.51 \\
\ion{Na}{i} D2& 1210.4237 & 4 & 120 &  99 & 0.05 \\
H$\beta$      & 1210.4237 & 7 &  10 &  53 & 0.22 \\
H$\gamma$     & 1210.4237 & 7 &  20 &  26 & 0.07 \\
\ion{Ca}{ii} K& 1210.4237 & 0 & --1 &  18 & 0.15 \\
\ion{Ca}{ii} H& 1210.4237 & 0 & --2 &  12 & 0.13 \\
\ion{Na}{i} D2& 1210.4237 & 0 &   2 &  13 & 0.61 \\
\ion{Na}{i} D1& 1210.4237 & 0 &   3 &  13 & 0.50 \\
H$\beta$      & 1210.5156 & 4 & 124 & 217 & 0.62 \\
H$\gamma$     & 1210.5156 & 4 & 117 & 218 & 0.51 \\
H$\delta$     & 1210.5156 & 4 & 111 & 223 & 0.40 \\
H$\epsilon$   & 1210.5156 & 4 &  76 & 262 & 0.60 \\
H$\zeta$      & 1210.5156 & 4 & 111 & 237 & 0.25 \\
\ion{Ca}{ii} K& 1210.5156 & 4 & 111 & 265 & 0.44 \\
\ion{Na}{i} D2& 1210.5156 & 4 & 132 & 111 & 0.04 \\
H$\beta$      & 1210.5156 & 7 &  10 &  53 & 0.23 \\
H$\gamma$     & 1210.5156 & 7 &  13 &  49 & 0.15 \\
\ion{Ca}{ii} K& 1210.5156 & 7 &  22 &  17 & 0.11 \\
\ion{Ca}{ii} H& 1210.5156 & 7 &  17 &  15 & 0.07 \\
\ion{Ca}{ii} K& 1210.5156 & 0 & --1 &  22 & 0.20 \\
\ion{Ca}{ii} H& 1210.5156 & 0 & --3 &  15 & 0.20 \\
\ion{Na}{i} D2& 1210.5156 & 0 &   3 &  13 & 0.61 \\
\ion{Na}{i} D1& 1210.5156 & 0 &   3 &  13 & 0.50 \\
\hline
\end{tabular}}
\end{table*}

\section{Results}

Table~\ref{master_table}  gives the radial velocity shift $v$, the velocity dispersion $\Delta v$ and the  absorption strength $R$ of each identified transient absorption, computed according to the procedure described in the previous section.
Uncertainties affecting the values of Table~\ref{master_table} are difficult to quantify, especially in the case of blended components.
We are confident, however, that the values are a good representation of the gas properties in the case of sharp, isolated features.
The results allow us to identify trends, based on the similar and consistent behaviour of many different lines.

\begin{figure*}
\centerline{ \
\includegraphics[width=0.80\hsize,angle=180,clip=true]{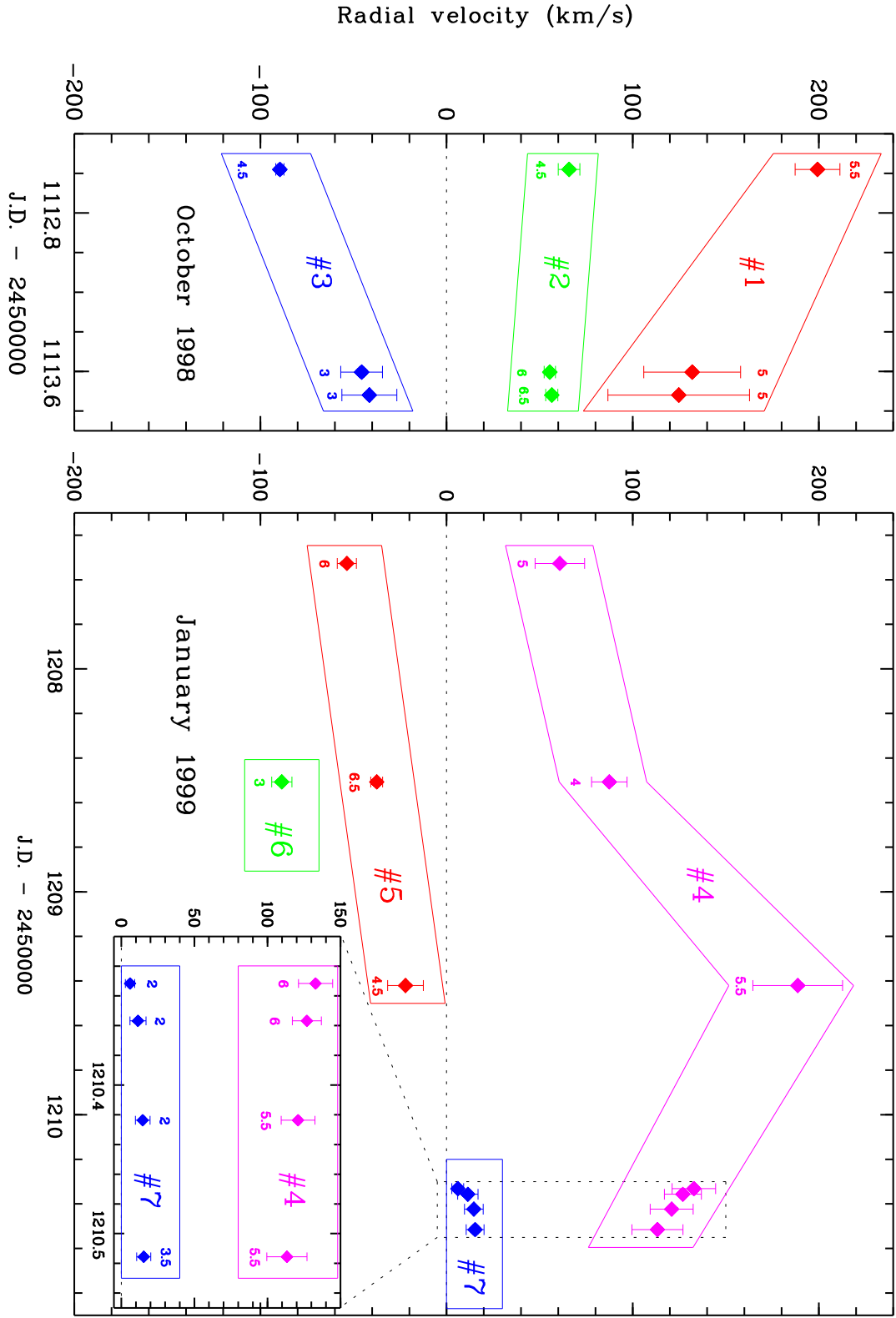}}
\caption{
Observed transient absorption components in each of the UES spectra (characterized by the Julian date of the observations).
Each point corresponds to one ``TAC'' and represents the mean velocity of the features with similar radial velocities detected in different lines. 
The radial velocity behaviour allows us to follow the temporal evolution of the gas causing the ``TAC"".
7 different dynamic events are identified; each event is enclosed in a box with an identification number (\#1, \#2,....).
Error bars show the rms error of the average velocity and the number of lines used to estimate the average velocity is indicated.
Fractional numbers come from the weight atributed to individual lines (see text).
For the sake of clarity, the spectra taken during January 31$^{\rm st}$ 1999 (JD 2451210.4) are expanded along the x--axis.
[{\it See the electronic edition for a colour version of this figure}] }
\label{master}
\end{figure*}

\begin{figure*}
\includegraphics[height=0.5\hsize,angle=-90,clip=true]{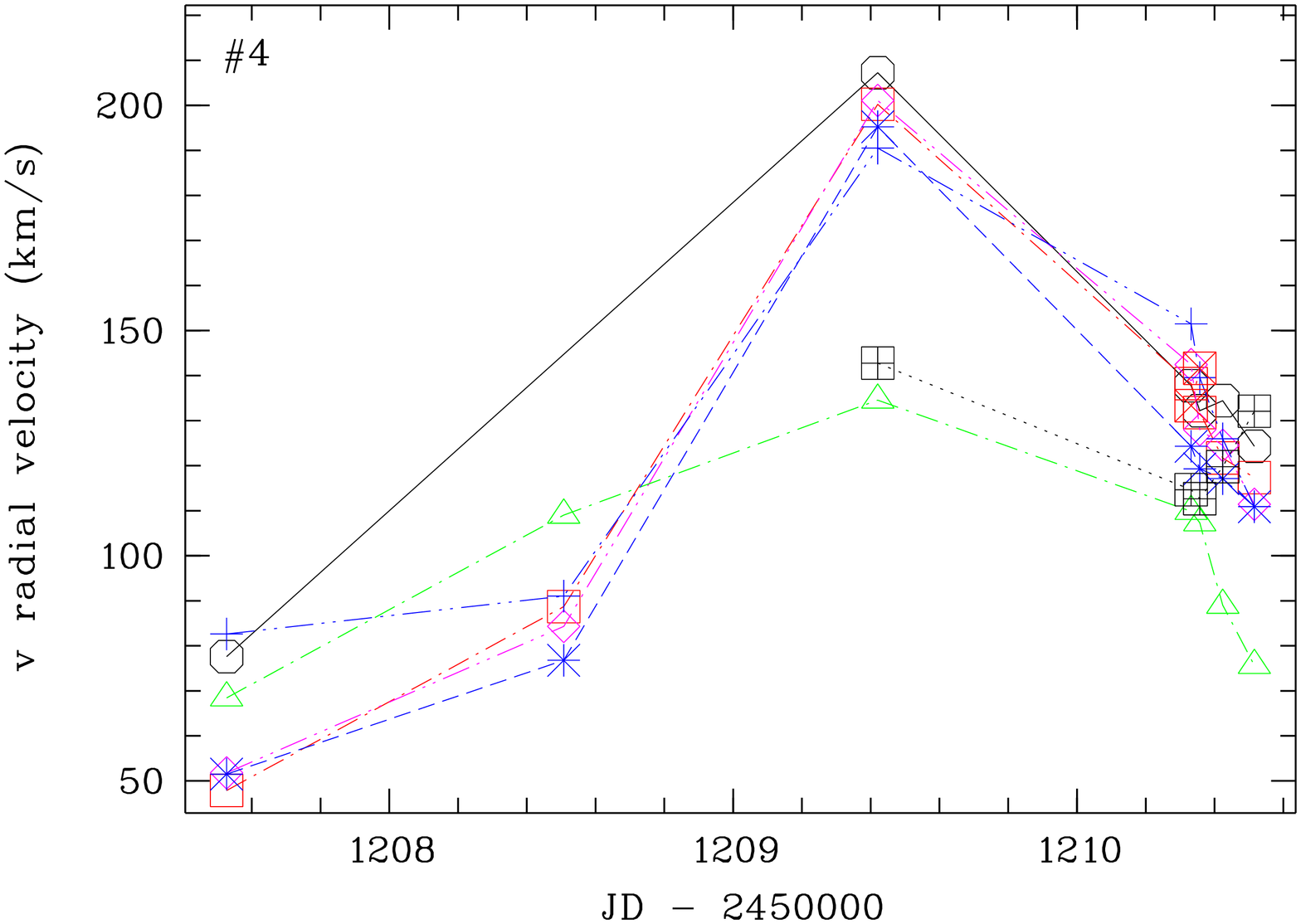}
\includegraphics[height=0.5\hsize,angle=-90,clip=true]{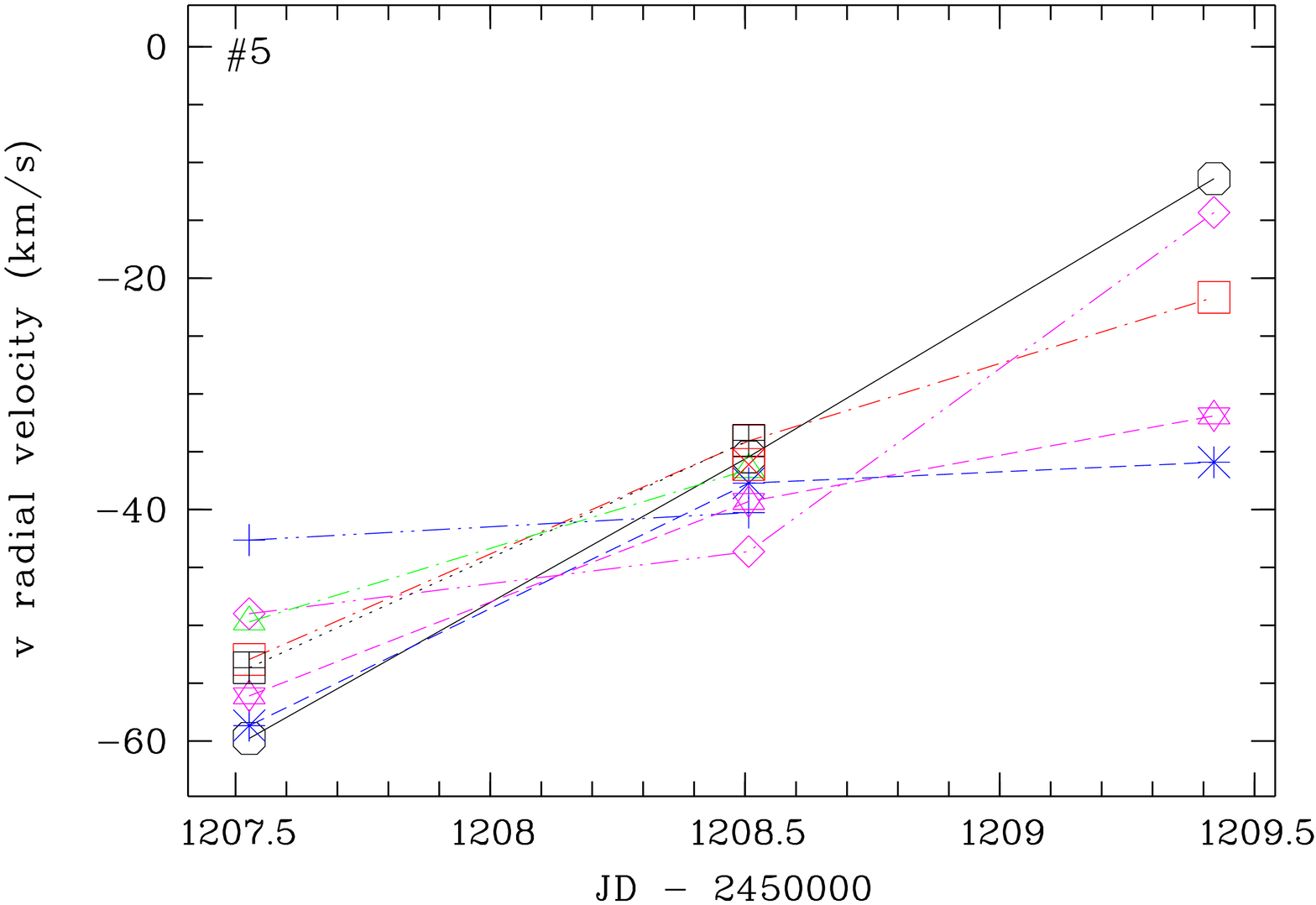}
\includegraphics[height=0.5\hsize,angle=-90,clip=true]{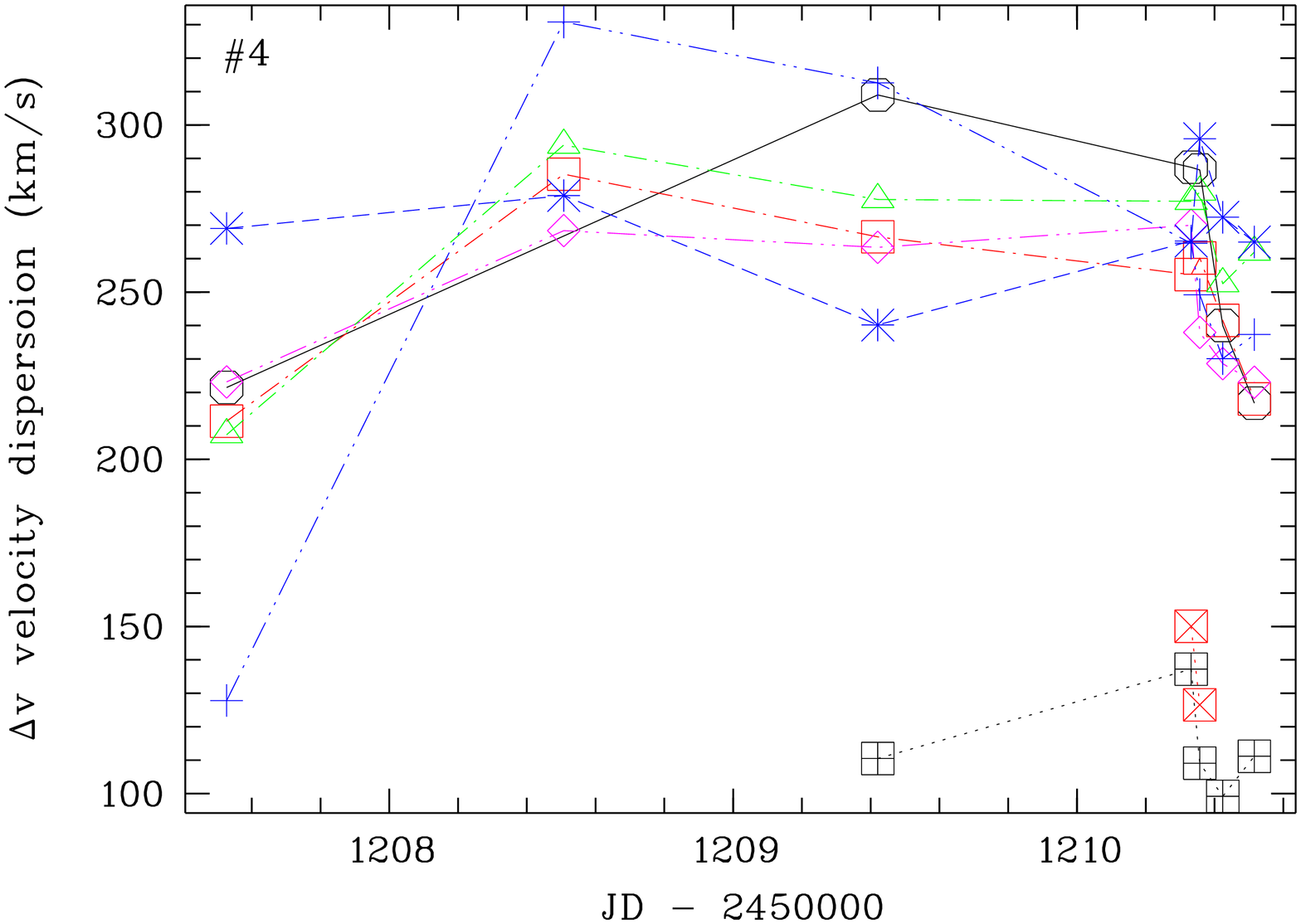}
\includegraphics[height=0.5\hsize,angle=-90,clip=true]{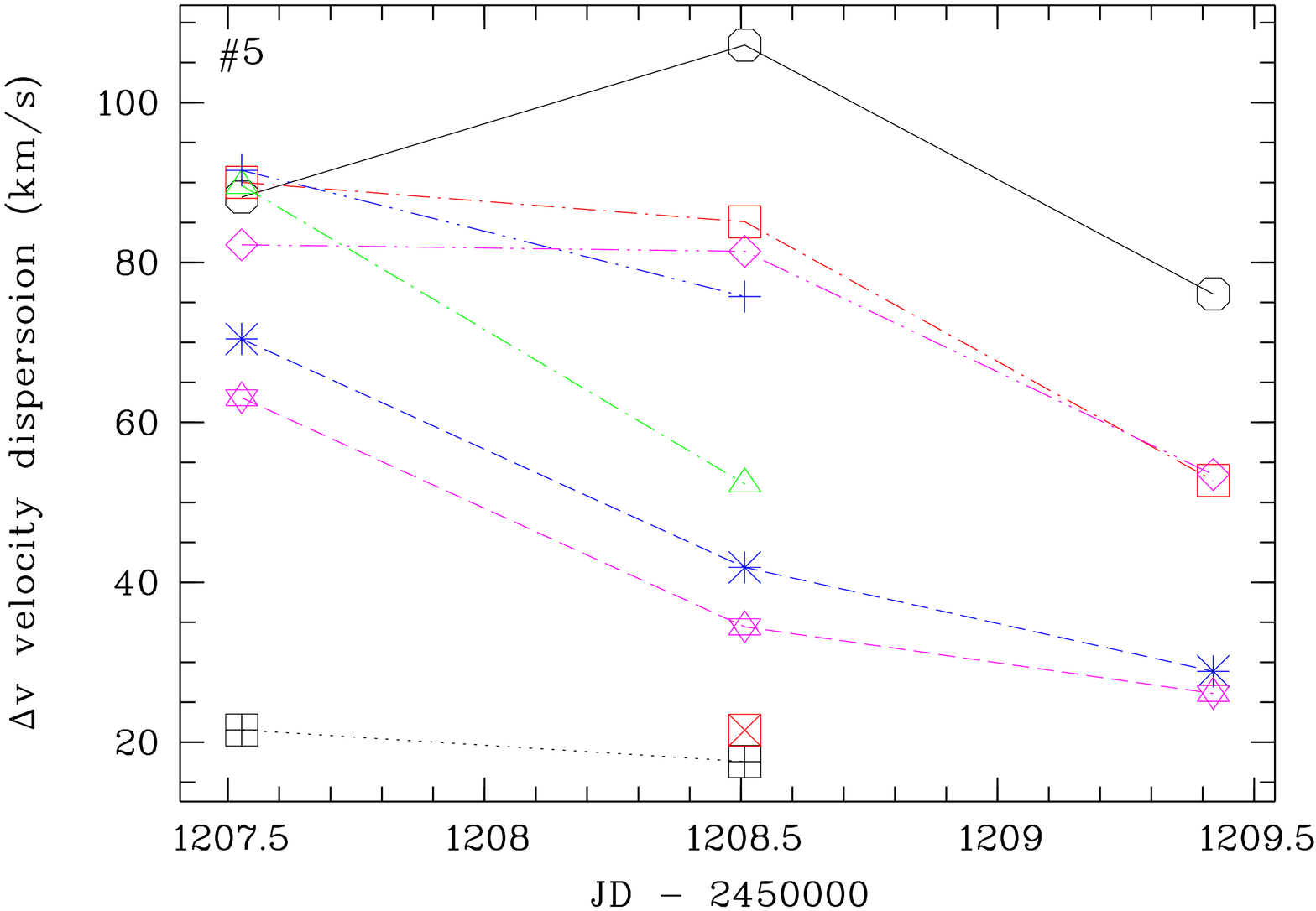}
\includegraphics[height=0.5\hsize,angle=-90,clip=true]{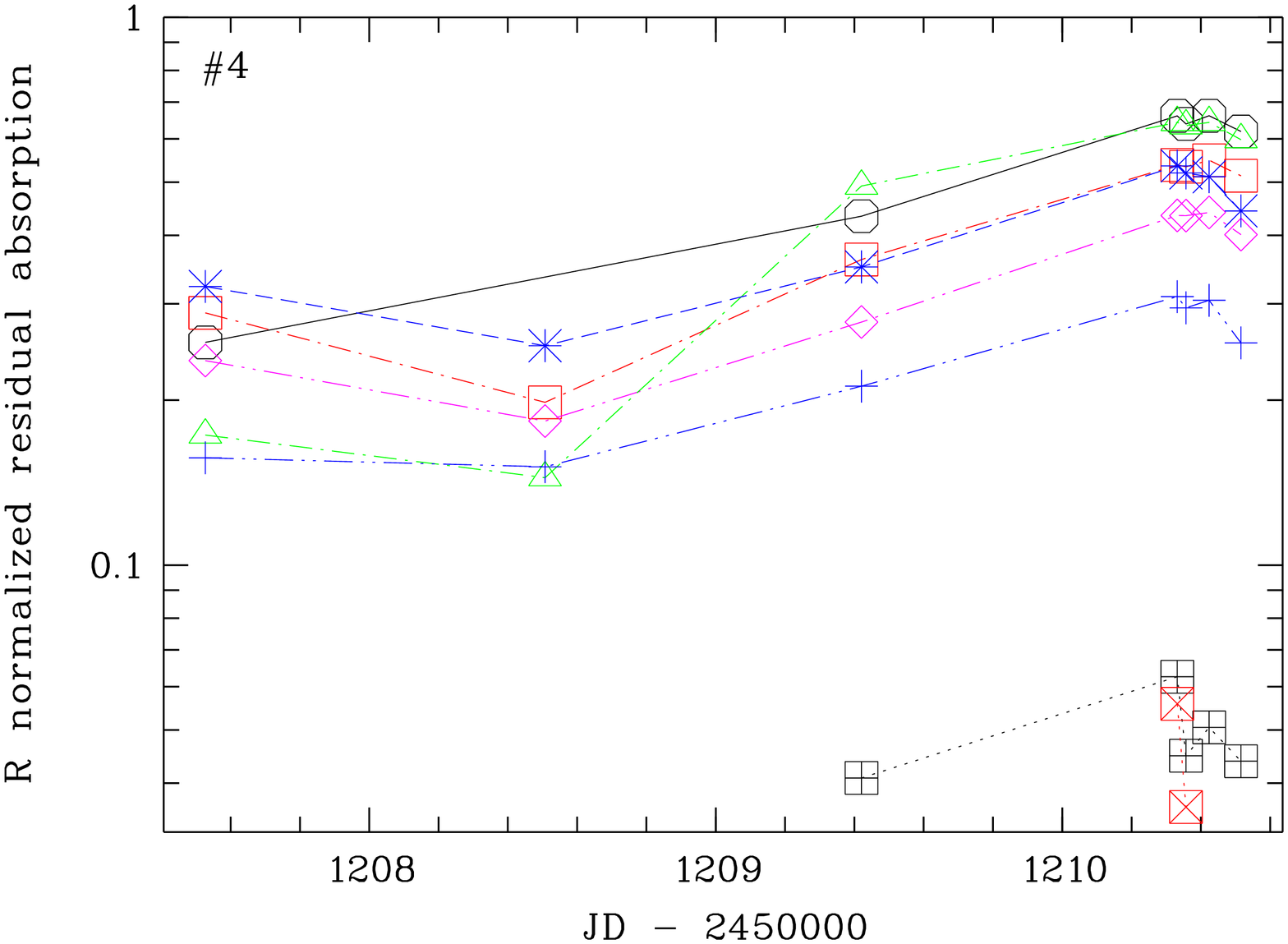}
\includegraphics[height=0.5\hsize,angle=-90,clip=true]{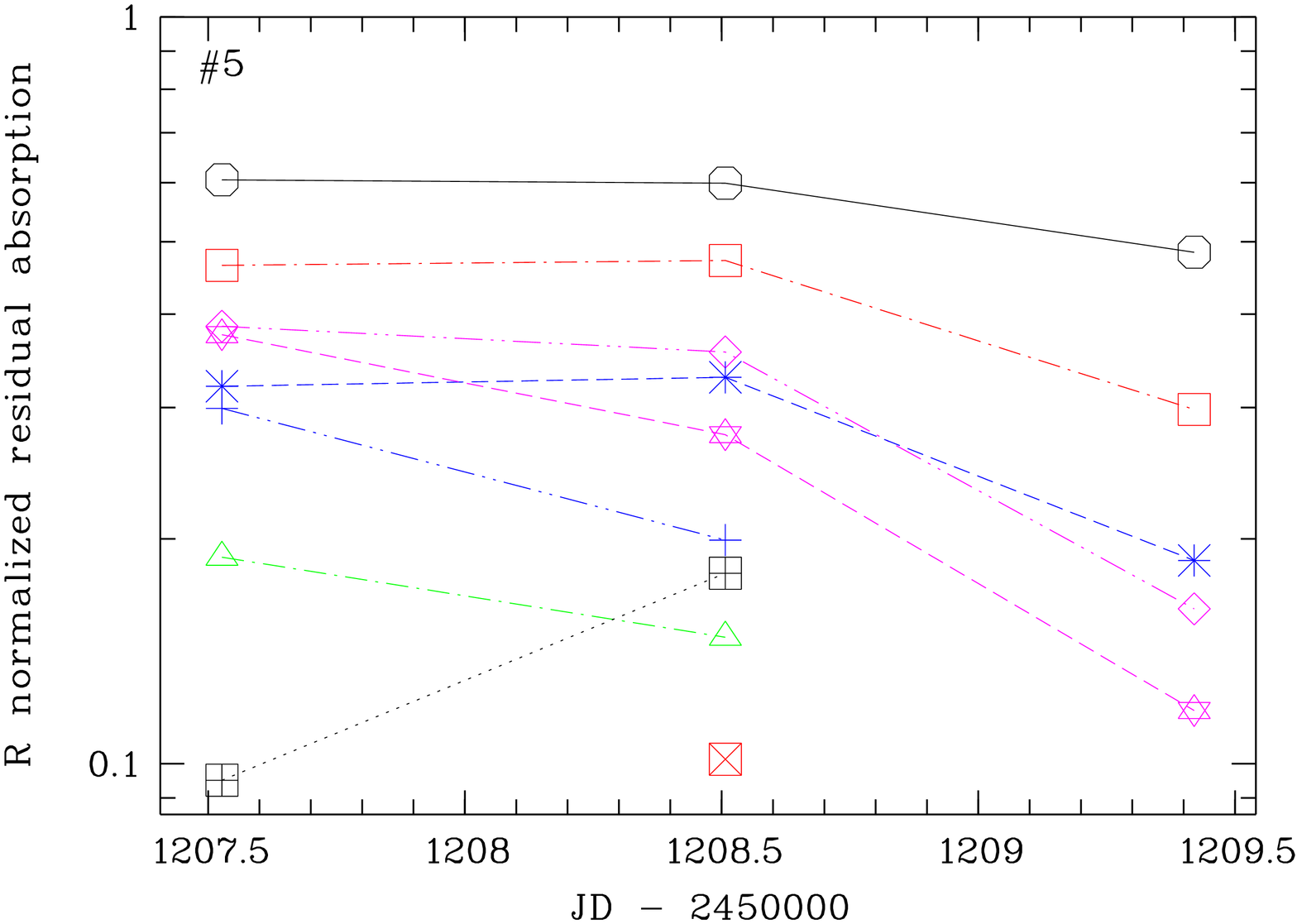}
\includegraphics[height=1.0\hsize,angle=-90,clip=true]{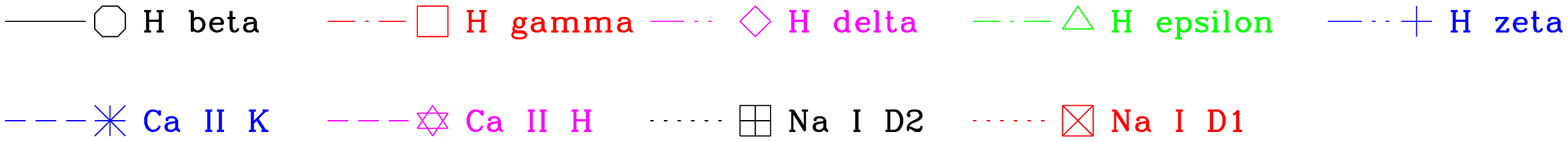}
\caption{Events \#4 and \#5.
The temporal evolution of $v$, $\Delta v$ and R is shown for every line as a function of the JD.
From left to right: events \#4 and \#5.
From top to bottom: time evolution of $v$ (km/s), $\Delta v$ (km/s) and R.
[{\it See the electronic edition for a colour version of this figure}] }
\label{events_4_5}
\end{figure*}

Absorption components at different radial velocities - either redshifted or blueshifted - are found in the Balmer and metallic lines. 
\ion{Ca}{ii} and \ion{Na}{i} components with the stellar radial velocity are also detected.
Features in different lines with similar radial velocities are detected within each spectrum; thus,  it is reasonable to assume that they form in approximately the same region.
The absorptions with similar velocities appearing in different lines are called a TAC, which can be characterized by the average of the radial velocity of the individual lines.
A total of 24 TACs can be identified in our spectra - 17 RACs and 7 BACs.
Fig.~\ref{master} plots the average radial velocity of each detected TAC as a function of the observing time.
Error bars show the rms error and the number of individual lines used to estimate the average velocity is indicated.
Fractional numbers come from the fact that a weight of 1/2 is assigned to those absorption line features that are less certain (the blended lines H$\epsilon$ and \ion{Ca}{ii}~H, the weak line H$\zeta$ and the \ion{Na}{i} doublet affected by telluric lines), and weight 1 to the rest.
Fig.~\ref{master} clearly shows that more than one TAC is present in each observed spectrum and that there is a radial velocity shift when comparing TACs in different spectra. 
The shift does not seem to be accidental; there appears to be a systematic temporal evolution of the TACs from one spectrum to the following one when their radial velocities are analysed.
This is clearly evident in the behaviour of the TACs observed in the four spectra taken during the night J.D.  2451210.5 with a time interval of around 4 hours from the first to the last one.  
Thus, the data reveal groups of TACs representing a dynamical evolution of the gas with which they are associated.
We call each of these groups an "event". 
In total, 7 events are identified, of which 4 are redshifted and 3 blueshifted.
The detailed temporal evolution of $v$, \Dv\ and $R$ for each individual line absorption component of 2 of these events (1 blueshifted and 1 redshifted) is shown in Fig.\ref{events_4_5}.  

\subsection{Kinematics}

In this section we describe the kinematic behaviour of the circumstellar gas as revealed by our observations.
Since they cover only a limited amount of time, it is difficult to estimate if the trends we have observed are always present, or if they are specific to our observations, and a different choice of times, or a longer coverage of the star, would give a different picture.
Only additional observations, and longer time coverage, can tell us what is the case.
With this caveat, this database is nevertheless a good starting point for the discussion of the TAC phenomenon in UX~Ori.

Among the detected events, the infalling gas (RACs) shows the largest velocities.
For instance, event \#1 shows a maximum average radial velocity of about 200 km/s which is roughly a half of the value of the stellar escape velocity, $\sim$410 km/s, while the maximum average velocity of the outflowing gas (BACs) is about 100 km/s.
This observational result could be biased since we begin detecting outflowing gas at maximum velocities, which is not always the case for the infalling gas (e.g. events \#4 and \#7). 
We also note that blueshifted gas is detected when there are redshifted absorptions, but the opposite is not true; we do not know if there is a physical meaning behind this result. 

The dynamics of events \#1 and \#2 (RACs) and \#3 and \#5 (BACs) denotes a deceleration of the gas.
In event \#4 the gas first accelerates and then decelerates; the data of \#7 suggest an acceleration, while event \#6 is only present in one spectrum.
An estimate of the acceleration of the events over the time interval covered by the data is given in Table~\ref{acceleration}.
Positive values mean acceleration while negative ones mean deceleration.
Table~\ref{acceleration} also gives the duration of each event, which is defined as the time it takes to reach zero velocity (for decelerating gas) or to go from zero to the maximum observed velocity for accelerating gas.
The estimates suggest that the events typically last for a few days and that accelerations are a fraction of \ms.
It is interesting to point out that the acceleration and deceleration phases of event \#4 are of the same order of magnitude.
We also note that our data do not allow us to conclude that this event is unique, since we may have missed the accelerating part of the other events.
In this respect, it would have been of great interest to follow the evolution of event \#7, which represents infalling gas with a significant acceleration, departing from a velocity very close to the stellar velocity (though it is only well detected in H$\beta$ and H$\gamma$).

\begin{figure*}
\centerline{\includegraphics[width=0.6\hsize,clip=true,angle=-90]{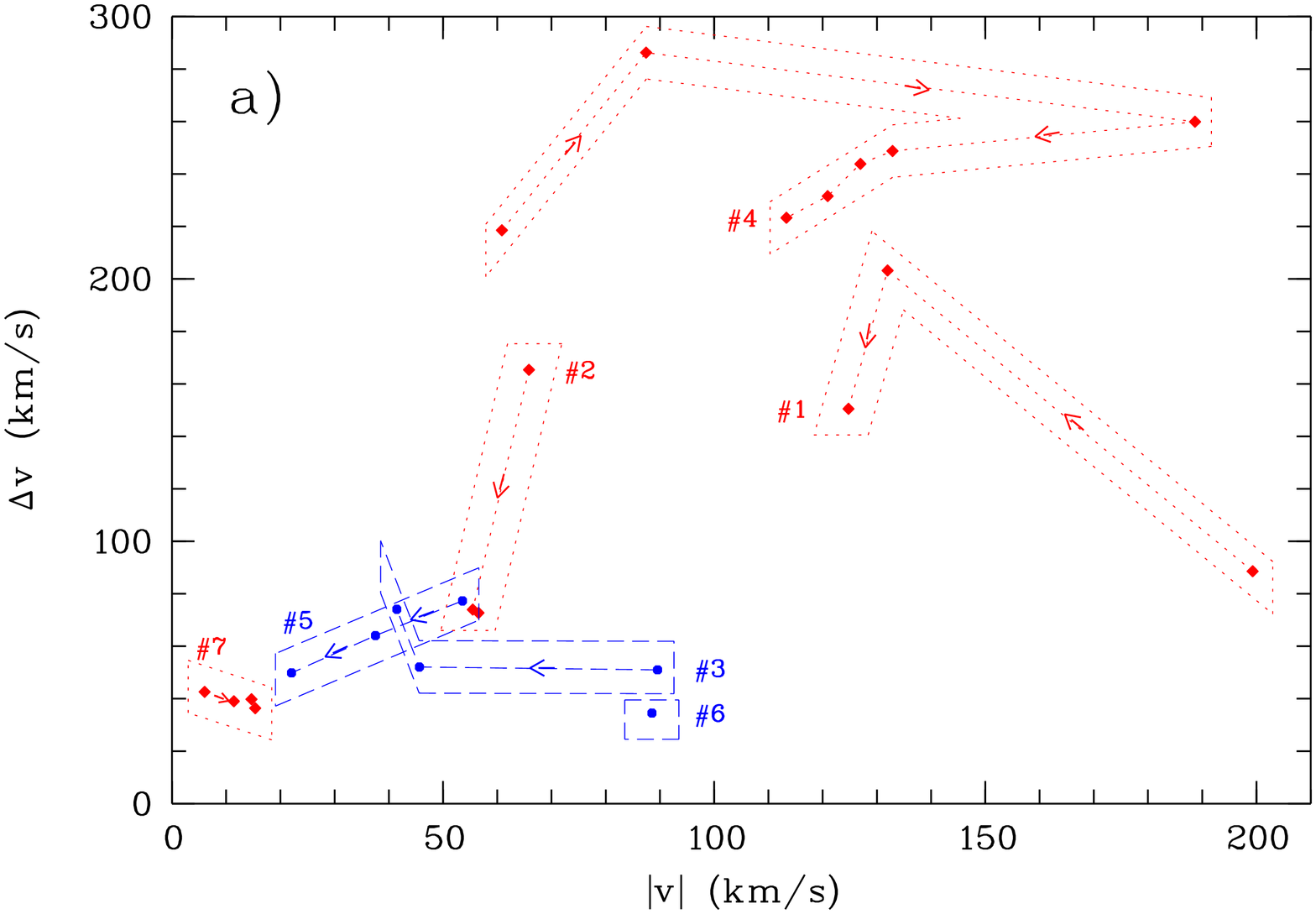}}
\centerline{\includegraphics[width=0.6\hsize,clip=true,angle=-90]{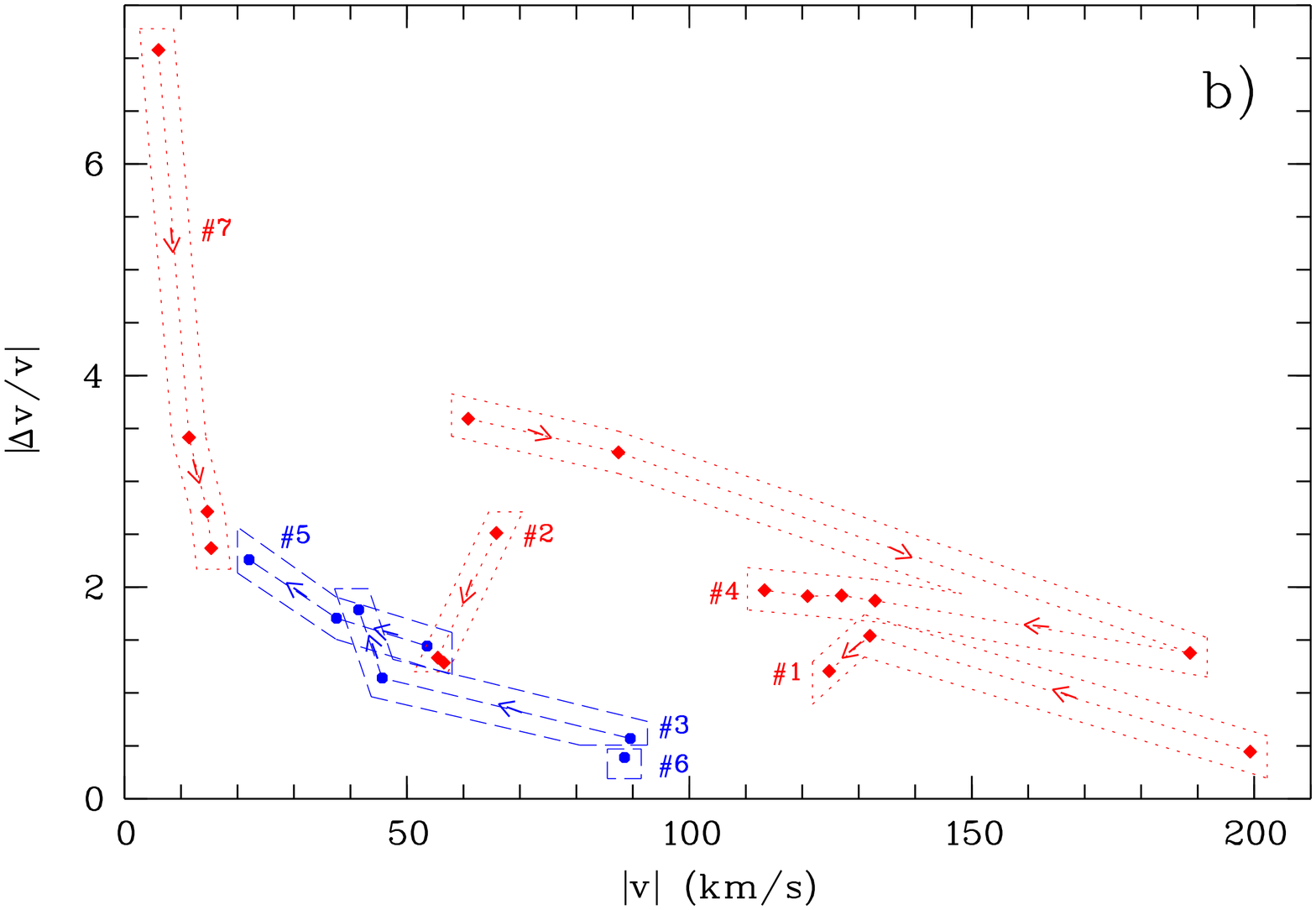}}
\caption {\Dv\ versus $v$ (top) and the complementary $|\Delta v/v|$ versus $v$ diagrams of all detected TACs.
Events enclosed in boxes with dotted lines are infalls and these in dashed lines are outflows.
The arrows show the dynamical evolution (acceleration/deceleration) of the events.
Redshifted gas appears to be broader than blueshifted gas.
The location of each event in the diagrams suggests that its velocity dispersion does not change substantially throughout the dynamical lifetime.
[{\it See the electronic edition for a colour version of this figure}] }  
\label{deltav_vs_v}
\end{figure*}

All the transient line absorption features are very broad (Table~\ref{master_table}), as is evident in the raw spectra. 
We have computed for each TAC of each event the average value of \Dv, weighted as in the case of $v$ (the rms dispersion of \Dv\ among the different lines in a TAC is typically $\sim$25\%).
There is a tendency for \ion{Na}{i} lines to be narrower than hydrogen and \ion{Ca}{ii} lines ($\Delta v_{\rm \ion{Na}{i}}~\simeq 0.5~\Delta v_{\rm H,\ion{Ca}{ii}}$).
This will not be investigated further in this paper, but it could give interesting clues to the physical conditions of the gas.
Fig.~\ref{deltav_vs_v} shows  \Dv\  versus $v$ and \Dv/$v$ versus $v$ for each detected TAC, which are two complementary views.
Each event with the corresponding TACs are enclosed in boxes.
The diagrams show that events are well separated and that the velocity dispersion does not change drastically along their temporal evolution (perhaps with the exception of event \#2). 
This behaviour is remarkable in event  \#4 which has both acceleration and deceleration phases.
Thus, these results suggest that events could be characterized by a kind of ``intrinsic'' velocity dispersion throughout their lifetimes.
 
\begin{table}
\caption{
Estimates of the acceleration and duration of each transient event.
The time scale for the event's lifetime is estimated as $\tau = |V_{\rm max}/a|$.
The acceleration and deceleration phases of event \#4 are shown separately.
$\tau$ for event \#7  means the time needed to achieve V$_{\rm max}$ departing from zero velocity. 
}
\label{acceleration}
\centerline{
\begin{tabular}{llll}
\hline
\hline
Event & Type     & $a$(\ms) &$\tau$ (days)\\
\hline
1     & infall   & -0.8  &  3.0           \\
2     & infall   & -0.1  &  8.1           \\
3     & outflow  & -0.5  &  2.1           \\
4$+$  & infall   & +0.8  &  2.6           \\
4$-$  & infall   & -0.8  &  2.7           \\
5     & outflow  & -0.2  &  3.2           \\
7     & infall   & +0.6  &  0.3           \\
\hline
\end{tabular}
}
\end{table}

\subsection{$R$ parameter}

Table~\ref{master_table} gives the parameter $R$ for each absorption component.
The ratio of the $R$ values among different lines hardly varies for all 24 identified TACs.
This is shown in Table~\ref{ratios_table}, where we give the ratio of \Hb, \Hd, \Hz, \ion{Ca}{ii}~K and \ion{Na}{i}~D2 to \Hg.
H$\gamma$ has been chosen as the reference line because it is clearly present in all events.
To estimate the errors we have applied a sigma-clipping algorithm to reject bad points; the corresponding values are given in Table~\ref{ratios_table}. 
The remaining lines do not have a well defined ratio to \Hg; we think this has  no physical meaning, since H$\epsilon$ and \ion{Ca}{ii}~H are blended and the relatively weak \ion{Na}{i}~D1 is more affected than \ion{Na}{i}~D2 by the strong telluric absorptions in this wavelength interval.

\begin{figure*}
\centerline{\includegraphics[height=\hsize,clip=true,angle=-90]{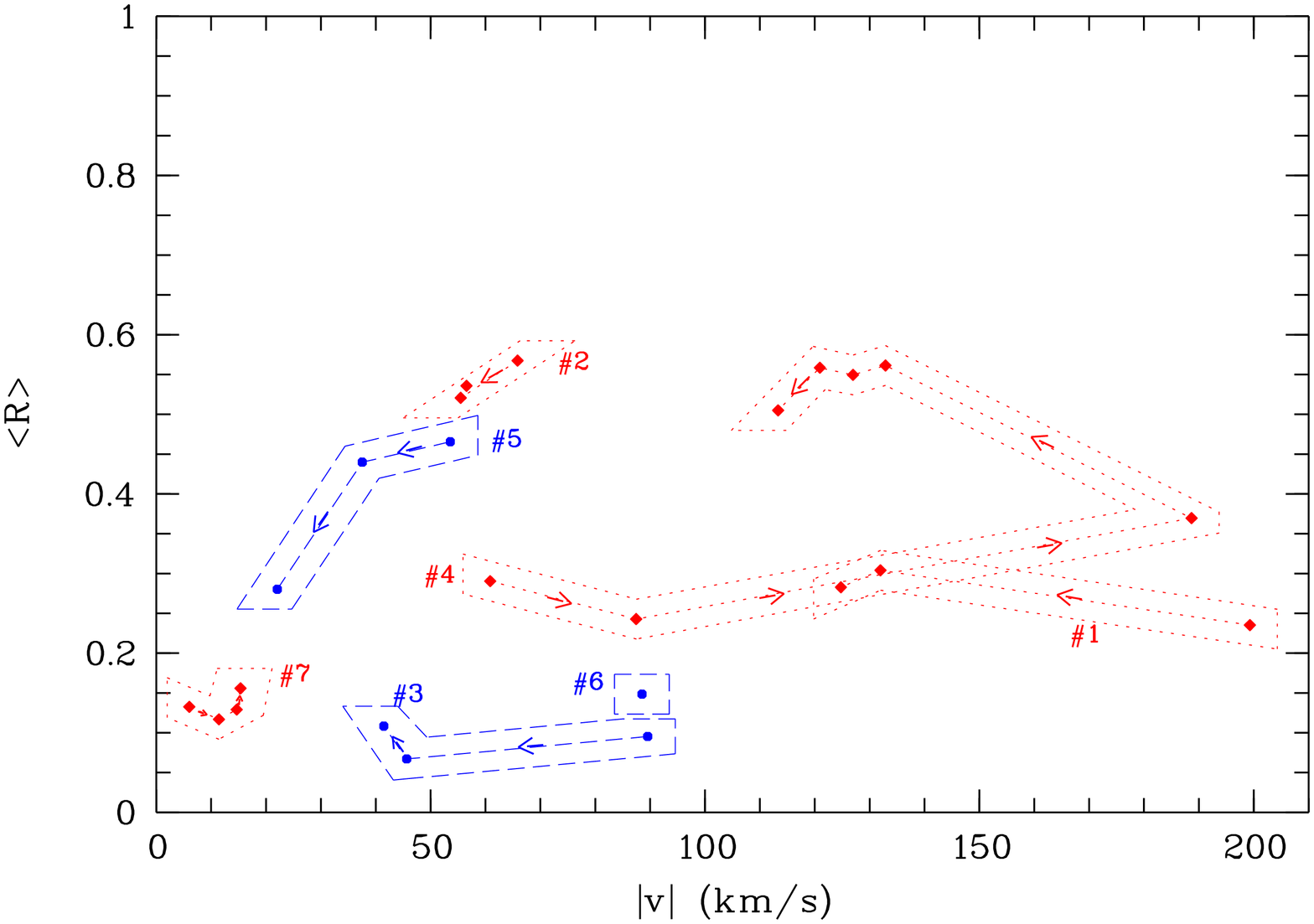}}
\caption {``Intensity'' $<R>$ versus radial velocity $v$ of each TAC.
Events and their dynamical evolution are marked as in Fig~\ref{deltav_vs_v}. 
$<R>$ is different for each event, but there seems to be no large changes among the TACs which form an event.
[{\it See the electronic edition for a colour version of this figure}] }  
\label{r_vs_v}
\end{figure*}

The fact that the ratio of $R$ between different lines is rather constant allows us to characterize each TAC by an ``intensity" ($<R>$) in the following manner.
The ratios $R_{\rm line}$/$R_{\rm H\gamma}$ of each TAC are used to compute an ``equivalent average $R_{\rm H\gamma}$ $<R>$". $<R>$ is estimated from H$\beta$, H$\delta$, \ion{Ca}{ii}~K (weight 1.0) and H$\zeta$ (weight 0.5).
Fig.~\ref{r_vs_v} shows $<R>$ as a function of the TAC velocities and events are enclosed in boxes.
No correlation is seen. 
However, as in the case of \Dv, the estimated $<R>$ values do not seem to change drastically from TAC to TAC within individual events.     

\begin{table}
\caption{Ratios of the average $R$ parameter of several lines to H$\gamma$.
The ratios (column 2) are estimated using all TACs in which the corresponding line absorption is present.
A sigma-clipping algorithm has been applied in order to reject bad $R$ values. 
The statistical error, the sigma-clipping value adopted and the fraction of the rejected values are given in columns 3 to 5.
The theoretical value for the Balmer lines is shown for comparison (gf$_{\rm line}$/gf$_{\rm H\gamma}$
).}
\label{ratios_table}
\centerline{
\begin{tabular}{llllll}
\hline
\hline
Line           & Ratio  & Error & clip. $\sigma$& \% Rej. & Theor. \\
\hline
H$\beta$       & 1.2    & 0.2   & 2.5           &  13     & 2.67   \\ 
H$\delta$      & 0.8    & 0.2   & 2.5           &  0      & 0.49   \\ 
H$\zeta$       & 0.6    & 0.15  & 2.5           &  0      & 0.18   \\ 
\ion{Ca}{ii} K & 0.9    & 0.2   & 2.5           &  5      &  --    \\
\ion{Na}{i} D2 & 0.14   & 0.06  & 2.0           &  20     &  --    \\
\hline
\end{tabular}
}
\end{table}

\section{Discussion}

The intensity and dynamics of the transient components are revealed by the spectra in such detail that a careful analysis in the context of  specific models (magnetospheric accretion models, to mention one obvious example) could be used to validate the theoretical scenarios.
This is in itself a huge effort, well beyond the scope of this paper. There are, however, several points that we think are worth mentioning.
Firstly, the data indicate that the events are the signatures of the dynamical evolution of gaseous clumps in the UX Ori CS disk.
There do not seem to be large changes in the velocity dispersion and the relative line absorption strengths through the lifetime of each clump.
This suggest that they are `blobs' which basically preserve their geometrical and physical identity.
In this respect, event \#4 is remarkable since during its estimated life of $\sim$6 days it shows both an acceleration and a deceleration.
In addition, the data reveal other aspects concerning the nature and dynamics of the events which are indicated in the following subsections.  
 
\subsection{Origin of the circumstellar gas}

Based on spectra qualitatively similar to those discussed here, \citet{natta2000} analysed the chemical composition of one infalling event in UX Ori.
Non-LTE models were used to estimate the ionization and excitation of Balmer (up to  H$\delta$), Ca II and NaI lines.
Roughly solar abundances were found for the gas causing the redshifted features of that event.
The present data suggest a similar nature for the gas causing the identified blobs.
We detect evidence of infalling gas in the Balmer lines and in a number of metallic lines with similar velocities, which also display roughly constant strength ratios to the Balmer lines; these ratios are $< 1$ (see Table~\ref{ratios_table}).
Similar result are found for the outflowing gas (blueshifted absorptions).
Thus, we can conclude that the CS gas in UX Ori is not very metal rich.
In addition, the data suggest that the physical conditions of the outflowing and infalling gas are rather similar and they likely co-exist in space. 

A second point is that in both RACs and BACs the CS gas has a significant underlying emission, at least in the Balmer lines we have examined.
This can be seen easily if we compare the values of the observed average ratio for the Balmer lines to the ratio of the opacity in the lines (see Table~\ref{ratios_table}, columns 2 and 6).
If the CS gas was just absorbing the photospheric flux, then the ratio of any two lines originating from the same lower level should be equal to the ratio of their opacities.
This is clearly not the case (see also similar results in \citet{natta2000}).
There are a number of reasons why this can happen.
The first is that the observed absorption is due to a very optically thick cloud, whose projected size is smaller than the stellar surface.
In this case, however, $R$ should be the same for all the lines, roughly equal to the occulted fraction of the stellar surface, and this does not seem to be the case.
In particular, the \ion{Na}{i}~D1 absorption is always weaker than the \ion{Na}{i}~D2 (R$_{\rm \ion{Na}{i}D2}$~/~R$_{\rm \ion{Na}{i}D1}$~=~1.4~$\pm$~0.4).
A second possibility is that there is significant emission in the lines.  

R, for any given component, can be estimated from the ratio $F_{\rm obs} / F_{\rm syn}$.
This can be done following eq.~5 of \citet{rodgers2002}, which provides this ratio under the assumption of a stellar and a circumstellar contribution to the observed flux; the circumstellar contribution is caused by a spherical occulting cloud and a more extended envelope.
In our zero order approach we neglect the extended envelope, and we assume a small optical depth and black-body emission for the spherical occulting cloud.
In this case, R can be written as:

\begin{equation}
R= \tau \>\>\Big[1- {{B_\nu(T_{\rm ex})}\over{B_\nu({T_\star})}} \> 
\big({{R_{\rm cloud}}\over{R_\star}}\big)^{2}\Big]
\end{equation}

where $\tau$ is the optical depth in the line, and $T_{\rm ex}$ is the black-body excitation temperature of the gas in the cloud. $R_{\rm cloud}$ is the projected radius of the CS cloud.
Some "filling-in" of the absorption features ($R < \tau$) may occur if $T_{\rm ex} \stackrel{<}{\sim} T_\star$ and $R_{\rm cloud} \stackrel{>}{\sim} R_\star$ (e.g. $T_{\rm ex} \simeq 7000 K$ and $R_{\rm cloud}/R_{\rm \star} = 1.6$, which is the corotation radius for UX~Ori, reproduce the observed ratios).

\subsection{Gas Dynamics}

As already pointed out, the discussion in the previous subsection together with the results of \citet{natta2000} indicates that the TACs in UX Ori do not arise from the evaporation of solid bodies, but in a substantially different scenario.  
The simultaneous presence of infalling and outflowing gas in the immediate vicinity of \UX\ is reminiscent of the predictions of magnetospheric accretion models, as developed by a number of authors in the last several years (see \citet{hartmann1998} for a basic account of this theory).
All these models assume stationary conditions, and it is difficult to derive quantitative predictions on the time dependent behaviour of the CS lines.
However, some simple points can be made to constrain the different models.

Let us start with the infall motions we detect. 
The simplest case we can consider for comparison is that of a blob of gas moving along the field lines of a stellar dipole magnetic field, seen in absorption against the stellar photosphere.
The infalling gas is practically contained in a region of the size of the corotation radius, which for \UX\ is $R_{\rm co} \sim 1.6$ \Rstar. 
This is in agreement with the discussion in subsection 5.1.
The small value of the corotation radius implies that the maximum poloidal velocity the blob will reach is of order $\sim V_{\rm esc} (1-R_{\rm star}/R_{\rm co}) \sim 220$ \kms.
 Since  the \UX\ disk is seen almost edge-on \citep{voshchinnikov1988}, the  observed velocity shift will be of the same order, which is in agreement with what we observe.
A simple calculation shows that the blob will be seen to accelerate from very low velocity to this maximum value in a time interval of about 1 day, until it hits the star.
The acceleration seen in event \#4 is reminiscent of such a behaviour, although the time scale is longer.
On the contrary, the decelerating infall motions we  see (events \#1,2,4) require a modification of this simple model.
A significant distortion of the velocity pattern can be caused by the rapid rotation of the star, which can create a sort of centrifugal barrier for the infalling gas.
Using the expression of the poloidal velocity given by eq. (8.76) of \citet{hartmann1998} for the \UX\ parameters, we find a  decrease of the maximum velocity (to about 180 \kms), but a similar acceleration pattern and time scale.
Still, the interaction of the fast rotating magnetosphere with the accreting disk material should be considered in detail.
\citet{muzerolle2001} have included rotation in a study of T~Tauri Stars (TTS), where, however, it is much smaller than the escape velocity at the stellar surface and the change of the poloidal velocity is negligible.
Similar calculations for cases where the stellar rotation is a significant fraction of the escape velocity  would be very valuable.

Outflowing gas is seen in events \#3 and \#5.
In both cases we start detecting the outflow at negative velocities of 100--80 \kms, and we see deceleration on time scale of 2--3 days.
Stellar-field-driven wind models \citep[X-wind, see ][ and references therein]{shu2000} predict that the wind leaves the disk surface close to the corotation radius on almost radial trajectories, accelerating rapidly to a speed of the order of $\Omega_\star R_{\rm co} \sim  350$ \kms.
This is significantly more than the maximum outflow velocity we observe, although the time coverage of our spectra can bias this result since we always find the maximum velocity of the outflow in the first spectrum taken in both the October 98 and January 99 observing runs.
Nevertheles, the deceleration we observe is not predicted by the models, unless collimation of the wind into the direction of the magnetic axis (i.e., away from the line of sight to the observer) occurs before the maximum speed is achieved and takes place on time scales of few days.
Winds driven by disk magnetic field \citep[see ][ and references therein]{konigl2000} tend to produce slower outflows with large velocity dispersions from a broad region of the disk.
\citet{goodson1997} discuss models where the interaction between the rotating stellar field and a disk field results in a time-dependent launching of a two component outflow, of which one (the disk wind) has some similarity to what we observe.
However, also in these models it is not clear what could cause the observed deceleration.
It is noteworthy that, once the acceleration phase is over, a wind can be decelerated by the effect of the stellar gravity \citep{mitskevich1993}. 
In \UX\, the deceleration caused by the central star is $\sim 0.2-0.5$ m\,s$^{-2}$ at distances of $\sim 14-20$ \Rstar, where the local escape velocity is $\sim$ 130 \kms.
An episode of outflow consistent with being gravitationally decelerated was observed in the TTS SU~Aur by \citet{petrov1996}.

The predictions of simple magnetospheric accretion models have been compared to the observed line variability in a number of TTS, with moderate success.
The best case is that of SU~Aur, a rather massive TTS with a rotation period of about 3 days.
This star shows evidence of simultaneous infall and outflow motions in various lines, similar to those observed in \UX\ \citep{johns1995,petrov1996,oliveira2000}.
The infalling velocity is clearly modulated by the rotation period of the star, and a model with magnetically chanelled accretion in a dipole field inclined by a few degrees to the rotation axis of the star may account for most of the observed infall properties.
As in \UX\, the outflow is  very variable, with maximum velocity similar to that of the infalling gas and a tendency to decelerate on a time scale of days, with no obvious rotational modulation \citep{johns1995}.
Note that the timing strategy of our \UX\ observations is not suited to the detection of modulations on the time scale of the stellar rotation period, which is about 17 hr.

\UX\ may provide a useful test-case.
On one hand, it is likely that the structure of the stellar magnetic field is more complex than in TTS, since \UX\ (a A-type star of 2--3 \Msun) lacks the surface convective layer that may create a simple magnetic field in TTS. 
On the other hand, one may take advantage of the fact that the average accretion rate of \UX\ is rather low \citep{tambovtseva2001}.
The CS gas is  mostly  seen in absorption against the continuum, when a sporadic increase of the accretion rate in the disk creates gas ``blobs" which move along the magnetic field lines, tracing their pattern more clearly than in objects with higher accretion rates, where broad emission dominates the CS lines.
We would like to mention that \citet{pontefract2000} also suggest magnetospheric accretion as a promising model to explain the H$\alpha$ spectropolarimetry data of the Herbig Ae star AB~Aur.

One final point to keep in mind in this context is that variations along the flow (both for infall and outflow) of the source function in the various lines may have an important effect on the comparison between observations and models, so that detailed calculations are required for validation of the models. 

\section{Conclusions}

The data presented in this paper allow us to analyse the spectroscopic behaviour of the CS gas disk around UX Ori on time scales of months, days and hours.
Significant activity in the CS disk is always present, which manifests itself in the continuous appearance and disappearance of absorption components detected in hydrogen and in many metallic lines.
This activity is not related to substantial variations of the stellar photosphere.
Blobs of gas experiencing infalling and outflowing motions are the likely origin of the transient features.
Blobs undergo accelerations/decelerations of the order of tenths of m\,s$^{-2}$ and last for a few days.
Detectable changes in the gas dynamics occur on a time scale of hours, but the intrinsic velocity dispersion of the blobs appears to remain rather constant.
No noticeable differences are seen in the properties of the infalling and outflowing gas, although infalls might have larger velocity dispersion.   
The relative absorption strength of the transient absorptions suggests gas abundances similar to the solar metallicity, ruling out the evaporation of solid bodies as the physical origin of the spectroscopic features. 
We suggest that the data should be analysed in the context of detailed magnetospheric accretion models, similar to those used for T Tauri stars. 

\begin{acknowledgements}
The authors wish to thank V.~P.~Grinin for valuable discussion about the analysis procedures. 
A. Mora acknowledges the hospitality and support of the Osservatorio Astrofisico di Arcetri for two long stays during which a substantial part of this work was carried out.
A.~Alberdi, C.~Eiroa, B.~Mer\'{\i}n, B.~Montesinos, A.~Mora, J.~Palacios and E.~Solano have been partly supported by Spanish grants ESP98-1339 and AYA2001-1124.
\end{acknowledgements}

\bibliographystyle{aa}
\bibliography{2676}

\end{document}